\documentclass[onecolumn,natbib,lscape]{aa}
\usepackage{natbib}
\usepackage{graphicx}
\usepackage{graphics}
\usepackage{longtable}
\usepackage{lscape}
\usepackage{amssymb}
\usepackage{txfonts}
\bibstyle{aa} 
\def\arcsec{\ifmmode^{\prime\prime}\;\else$^{\prime\prime}\;$\fi}
\def\arcmin{\hbox{$^\prime$}}
\def\deg{\hbox{$^\circ$}}
\def\funits{\,{\rm erg ~cm^{-2} ~s^{-1}}}

\begin{document}

   \title{High redshift X--ray galaxy clusters. I. The impact of point 
          sources on the cluster properties}
    \subtitle{}
   \author{M. Branchesi
           \inst{1,2}, 
           I. M. Gioia
           \inst{2}, 
           C. Fanti
           \inst{2}, 
           R. Fanti
           \inst{2}}

   \offprints{M. Branchesi}

   \institute{Dipartimento di Astronomia, Universit\`a di Bologna,
              Via Ranzani 1, 40127 Bologna, Italy \\
                \email{m.branchesi@ira.inaf.it}
                           \and
            INAF - Istituto di Radioastronomia, Via Gobetti 101, 
            40129 Bologna, Italy \\
                \email{gioia@ira.inaf.it, 
                      rfanti@ira.inaf.it, cfanti@ira.inaf.it}}  

   \date{Received ......... 2007; accepted ......... 2007}

\abstract 
{The current generation of X-ray observatories like \textit{Chandra}
allows studies with very fine spatial details. It is now possible to
resolve X--ray point sources projected into the cluster diffuse
emission and exclude them from the analysis to estimate
the ``correct'' X--ray observables.}  
{We wish to verify the incidence of point sources on the cluster
thermal emission and to evaluate the impact of their non-thermal
emission on the determination of cluster properties.}
{To these ends we use a sample of 18 high-z (0.25$<$z$<$1.01) clusters
from the \textit{Chandra} archive and subtract the non-thermal
emission of the point sources from the extended thermal emission due
to the cluster itself. We perform a detailed analysis of the cluster
properties and compare the changes observed in the X--ray observables,
like temperature and luminosity or their inter-relation, when one
keeps the point sources in the analysis.}
{The point sources projected into the cluster extended emission
affect the estimates of cluster temperature or luminosity considerably
(up to 13\% and 17\% respectively). These percentages become even
larger for clusters with z$>$0.7 where temperature and luminosity
increase up to 24\% and 22\%, respectively.}  
{The conclusions are that point sources should be removed to correctly
estimate the cluster properties. However the inclusion of the point
sources does not impact significantly the slope and normalization of
the L$_{bol}$-T relationship since for each cluster the correction to
be applied to T and L$_{bol}$ produces a moderate shift in the
L$_{bol}$--T plane almost parallel to the best-fit of the ``correct''
L$_{bol}$--T relation.}


\authorrunning{Branchesi et al.}
\titlerunning{The impact of Chandra point sources on the cluster properties}

\maketitle

\section{Introduction}

\smallskip\noindent
X-ray studies of clusters of galaxies performed with X-ray telescopes
like {\em ASCA} \citep{Ta94}, {\em Beppo-SAX}
\citep{Pa97,Ma97,Bo97,Fr97} or {\em ROSAT} \citep{Tr84} have
attributed the total X-ray emission from clusters of galaxies to
thermal bremsstrahlung emission from the thin hot gas that fills the
regions between the cluster galaxies.  This is due to the optimal
consistence between the observed X-ray spectra and the expected 
thermal emission from the highly ionized hydrogen and helium in the 
intra-cluster medium at temperatures T $\sim 10^{7} - 1.5 \times 10^{8}$ K. 
The detection of point sources in the cluster fields has been hindered by the 
traditionally poor angular resolution of such X-ray telescopes.  
The current generation of X-ray observatories like \textit{Chandra} 
\citep{Van97} or XMM-{\em Newton} \citep{Sr01,Tur01} has revolutionized 
X-ray astronomy, enabling studies with very fine spatial details. In 
addition the improvement of detection techniques \citep[e.g. the 
multiscale wavelet detection by][]{Fre02} can now reliably separate 
small-scale source emission from surrounding larger-scale diffuse cluster
emission. Hence, it is now possible to subtract the non-thermal
emission of the point sources embedded in the cluster emission from
the extended thermal emission due to the cluster itself.
In recent years a number of detailed XMM-{\em Newton} and
\textit{Chandra} studies have allowed investigators to compute the
``correct X--ray observables'' and thus the ``correct scaling 
relations'' between them excluding the point sources from the
analysis of the cluster emission. 

\smallskip\noindent
In this paper we present a detailed analysis of the archival
{\em Chandra} data for a sample of high-z clusters. The aim is to
verify the incidence of point sources on the thermal emission of the
cluster and to evaluate their contribution to the cluster total emission.
In particular we will examine the point source effects on the cluster 
temperature and luminosity and on the L$_{X}$--T relation.
The sample analyzed consists of eighteen clusters with redshift in the
range 0.25 $<$ z $<$ 1.01, that was also used by \cite{Br07a} 
(from now on BR07) to check
for any overdensity of point sources in the inner region of clusters
of galaxies. In a companion paper (Paper II; Branchesi et al. 2007) we
will use these same clusters in combination with clusters taken from
the literature to revisit the ``correct L$_{X}$--T'' relation and check 
its evolution with redshift. 

\smallskip\noindent
All uncertainties in this work are at the 1 $\sigma$ confidence level, 
unless otherwise noted.  We use a $\Lambda$CDM cosmology with $\rm H_{0} = 70 
~km s^{-1} ~Mpc^{-1}$ and $\Omega_{\rm m} = 1-\Omega_{\Lambda} = 0.3$.
\tabcolsep 0.1cm
\begin{table}
\begin{center}
\caption{Cluster sample parameters and details of Chandra observations}
\scriptsize
\small
\begin{tabular}{lcccrccrc}
\hline 
\hline
Cluster name    &      z & RA     &      DEC  &Obs.ID& ACIS & Mode
& Exp. & ${\rm N_H}$\\ 
        &        & $hh ~mm ~ss$ & ~~\deg ~~~~\arcmin ~~~~\arcsec &   &  & & ks &
\small{$10^{20}$ cm$^{-2}$}\\ 
(1) & (2) & (3) & (4) & (5) & (6) & (7) & (8) & (9)\\
\hline
\hline
Abell\,2125           & 0.246& 15 41 12&   $+$66 16 01&  2207& I & VF & 79.7 & 2.77 \\
ZW\,CL\,1454.8$+$2233 & 0.258& 14 57 15&   $+$22 20 33&  4192& I & VF & 91.4 & 3.22 \\
MS\,1008.1$-$1224     & 0.302& 10 10 32&   $-$12 39 23&  926 & I & VF & 44.2 & 6.74 \\
ZW\,CL\,0024.0$+$1652 & 0.394& 00 26 35&   $+$17 09 39&  929 & S & VF & 36.7 & 4.19 \\
MS\,1621.5$+$2640     & 0.426& 16 23 36&   $+$26 34 21&  546 & I & F  & 30.0 & 3.59 \\  
RXJ\,1701.3$+$6414    & 0.453& 17 01 24&   $+$64 14 10&  547 & I & VF & 49.5 & 2.59 \\
CL\,1641$+$4001       & 0.464& 16 41 53&   $+$40 01 46&  3575& I & VF & 44.0 & 1.02 \\
V\,1524.6$+$0957      & 0.516& 15 24 40&   $+$09 57 48&  1664& I & VF & 49.9 & 2.92 \\
MS\,0451.6$-$0305     & 0.539& 04 54 12&   $-$03 00 53&  902 & S & F  & 41.5 & 5.18 \\
V\,1121$+$2327        & 0.562& 11 20 57&   $+$23 26 27&  1660& I & VF & 66.9 & 1.30 \\
MS\,2053.7$-$0449     & 0.583& 20 56 21&   $-$04 37 51&  1667& I & VF & 43.5 & 4.96 \\ 
V\,1221$+$4918        & 0.700& 12 21 26&   $+$49 18 30&  1662& I & VF & 79.4 & 1.44 \\
MS\,1137.5$+$6625     & 0.782& 11 40 22&   $+$66 08 18&  536 & I & VF & 117.5& 1.18 \\ 
RDCSJ\,1317$+$2911    & 0.805& 13 17 21&   $+$29 11 19&  2228& I & VF & 111.3& 1.04 \\  
RDCSJ\,1350$+$6007    & 0.805& 13 50 48&   $+$60 06 54&  2229& I & VF & 58.3 & 1.76 \\  
RXJ\,1716.4$+$6708    & 0.813& 17 16 49&   $+$67 08 26&  548 & I & F  & 51.5 & 3.71 \\
MS\,1054.4$-$0321     & 0.830& 10 56 59&   $-$03 37 37&  512 & S & F  & 67.5 & 3.67 \\  
WARPJ\,1415.1$+$3612  & 1.013& 14 15 11&   $+$36 12 00&  4163& I & VF & 89.2 & 1.10 \\
\hline
\label{tab1}
\end{tabular}
\normalsize
\end{center}
 \begin{itemize}
 \item [--] Column 1: Cluster name
 \item [--] Column 2: Spectroscopic redshift tabulated in the literature
 \item [--] Column 3-4: Right ascension and declination 
            (Equatorial J2000, HH MM SS.S, +DD MM SS.S) of the
            centroid of the {\em Chandra} photon distribution 
            in the 0.5--5 keV energy band assumed as the cluster center  
 \item [--] Column 5: Identification number of the observation
 \item [--] Column 6: Detector where the aimpoint lies (I, for  
           {\em ACIS-I} or S, for {\em ACIS-S}) 
 \item [--] Column 7: Observation mode (F for FAINT or VF for VFAINT)
 \item [--] Column 8: Exposure time in ks corresponding to the nominal
            exposure filtered to exclude time periods of high background
 \item [--] Column 9: Column density of Galactic hydrogen in units of
            10$^{20}$ cm$^{-2}$, obtained from the {\em Chandra} 
            X--ray Center (CXC) Proposal Planning Tool Colden (Galactic 
            Neutral Hydrogen Density Calculator):  NRAO-compilation by 
            \cite {Di90} 
\end{itemize}
\end{table}
\section{\textit{Chandra} data reduction and cluster analysis}

We extracted from the {\em Chandra} archive the X-ray data of 18 galaxy
clusters with redshift in the range 0.25$<$ z $<$1.01 and exposure
time greater than 30 ks.  Data reduction was performed using version
3.2.1 of the CIAO software ({\it Chandra} Interactive Analysis of
Observations; see web page {\em http://cxc.harvard.edu/ciao/index.html)} 
and version 3.0.3 of the CALDB (Calibration Database).  The sample is the 
same used by BR07 where the selection criteria and the data reduction
are described in more detail. For clarification purposes we report here
Table~\ref{tab1} of BR07 that lists the sample parameters and
the details of the {\em Chandra} observations. Clusters are
arranged in increasing redshift order.

\smallskip\noindent
The clusters of galaxies analyzed here appear as extended sources in
the \textit{Chandra} images. Differently from lower resolution
instruments, \textit{Chandra} is able to reveal point sources
overimposed but not necessarily associated with the pointed cluster.
Although such objects may be scientifically interesting in their own
right (see discussion in BR07) investigators can now filter
out any point source before fitting spectral models to the extended
emission.  For a correct analysis of the cluster properties the point 
source (non-thermal) emission should be subtracted from the cluster 
extended emission. Since one of the goals of our analysis is
to estimate the impact of these point sources on the cluster
properties,  two approaches have been adopted:
\begin{itemize}
\item Point sources are identified using the CIAO Detect package
WAVDETECT, which has the ability to work in complex fields with both
point and extended sources (see BR07 for details on the procedure
followed). The point sources are later removed from the event
file using the \textit{dmcopy} command.
\item Point sources are {\it not} removed from the event file in an 
attempt to evaluate their effects on cluster parameters.  
\end{itemize}
\subsection{Background subtraction}

The issue of background subtraction was carefully considered. For both
spatial and spectral analyses, it is necessary to associate a
background to the source events. For most of the clusters the
background was measured locally, within the same target field, in a
region free of point sources and in the vicinity of the cluster but
not contaminated by the cluster emission. We checked that variations of
the background intensity across the chip do not affect the background
subtraction by comparing the count rates in the cluster and in the
background regions at energies larger than 8 keV, where the signal
from the cluster should be nil.

\medskip\noindent
We also used the ``blank-sky'' data sets, following the steps and
recommendations of the Markevitch's COOKBOOK
\footnote{See {\em http://cxc.harvard.edu/contrib/maxim/acisbg/data/README} and\\ 
{\em http://cxc.harvard.edu/contrib/maxim/acisbg/COOKBOOK}}.  The
``blank-sky" background files were first reprocessed and reprojected
to match the corresponding cluster observation gain and position.
The background files were then normalized to the shorter exposure time
of the cluster observations. The background events were extracted from
the same region of the chip as the cluster to model spatial
variations in the background. Obtaining background spectra from
blank-sky data sets has the advantage to use the same region as the
source, thus eliminating potential systematic errors caused by spatial
variations of both the energy response and the effective area across
the chip. The method remains, however, vulnerable to temporal
variations in the spectrum of the particle background and also cannot
easily account for the strong directional variation of the Galactic
soft X-ray emission.

\medskip\noindent
We performed consistency checks for several clusters of the sample.
The properties obtained using background files derived from different
regions of the target field and from the ``blank-sky" data sets are
generally independent of the background used.  For the two most nearby
clusters of the sample, ZW\,CL\,1454.8$+$2233 and Abell\,2125, which
cover a very extended region, we considered more appropriate to
extract the background spectrum file from the ``blank-sky" data-sets.
For all the other clusters the background spectrum file was extracted
from regions within the same target field.

\subsection{Spatial analysis}
\subsubsection{Cluster surface brightness}
\label{spatial}
The spatial analysis of the cluster X-ray emission was performed
within images (with point sources excluded) in the energy range
0.5--5.0 keV. Under the assumption that the cluster gas is a symmetric
isothermal sphere with a density profile described by a $\beta$-model
\citep {Ca76}:
\begin{eqnarray} 
\rm n_{gas}(r) = n_{0,gas} \left(1+ \left(\frac{R}{R_{c}} \right)^2\right)^{-3\beta/2},
\label{cl:eqn2}
\end{eqnarray}
the radial surface brightness was modelled accordingly as:
\begin{eqnarray} 
\rm S(r) = S_{0} \left(1+ \left(\frac{R}{R_{c}} \right)^2 \right)^{-3\beta+1/2}
\end{eqnarray}
where $\rm R_{c}$ is the core radius and $\beta$ describes the slope of the 
density profile at large radii.

\smallskip\noindent 
For each cluster the $\beta$-model fit to the surface brightness was
performed using the exposure-corrected image with a constant background
included in the fit.  The best-fit parameters for $R_{c}$ and $\beta$
are listed in Table \ref{tab2}.

\smallskip\noindent 
Even if the simple $\beta$-model is known to poorly describe the
radial profile of \textit{Chandra} highly resolved clusters
\citep[see][]{Et04}, this model provides a good description of the
cluster for our pourpose, that is to extrapolate luminosities to
larger radii (see Sect.~\ref{comparison} and Paper II of Branchesi et
al. 2007). Exceptions are MS\,1054.4$-$0321 and
ZW\,CL\,1454.8$+$2233. The former cluster presents a clear western
substructure. The fit to the whole cluster provides unreasonably large
values for $\beta$ and $R_{c}$ \citep[as previously noted
by][]{Je01}. The fit improves once we mask the western substructure
from the main body of the cluster.  The values of $R_{c}$ and $\beta$
in Table~\ref{tab2} for MS\,1054.4$-$0321 were obtained in this
way. For ZW\,CL\,1454.8$+$2233 a single $\beta$-model is not an
acceptable representation of the radial profile (see Notes on
individual clusters in Sect.~\ref{notes}). Therefore no values of
$R_{c}$ and $\beta$ are indicated in Table~\ref{tab2}.

\subsubsection{Definition of cluster extent}
\label{radius}

The spectrum of each cluster was extracted from a circular region
centered on the X-ray centroid, out to a maximum radius which
is hereafter referred to as spectral radius, $\rm R_{spec}$.  This
radius was chosen individually cluster by cluster so as to optimize
the signal-to-noise ratio in order to determine the X-ray temperature 
of the cluster with maximum count statistics. To define this circular 
region around the centroid of the photon distribution we followed 
the procedure suggested by \cite{To03}.

\smallskip\noindent 
A second extraction radius was adopted in order to consider the
cluster extent where diffuse emission is still detectable.  This
radius, named $\rm R_{ext}$, is the radius  where the cluster radial
surface brightness merges into the background, and beyond
which no further significant cluster emission is detected.  Both $\rm
R_{spec}$ and $\rm R_{ext}$ are listed in Table \ref{tab2}.  The
fraction of the net counts included in the $\rm R_{spec}$ extraction
region is always between 0.80 and 0.95 of the net counts included in
the $\rm R_{ext}$ extraction region once point sources have been
excised from the cluster emission.

\smallskip\noindent 
For each cluster, the events included in the extraction region were
used to produce a spectrum file. This procedure was repeated
considering the two above defined radii ($\rm R_{spec}$ and $\rm
R_{ext}$) and approaches, that is filtering out or not the events
associated to the point sources from the event file. We ended up with
four spectrum files per cluster.

\subsection{Spectral analysis}
\label{fit}

Spectra were extracted from within both the defined detection radii, 
$\rm R_{ spec}$ and $\rm R_{ ext}$, considering the two situations
described above: cluster emission alone (from now on ``cluster'') and
cluster plus point source emission (from now on ``cluster$+$ps''). 
The Auxiliary Response File (ARF) and the Redistribution Matrix File (RMF)
were computed from the same region where the spectra are
extracted and were weighted by the detected counts in the restricted
energy range (0.5--2.0 keV) where both the response and the
thermal model do not vary much. When possible the weighted RMFs
have been created using the new CIAO tool \textit{mkacisrmf}\footnote{{\em
mkacisrmf} has been calibrated for the {\em ACIS-I} array plus {\em
ACIS-S1,S2}, and {\em S3}.  The tools creates response files intended
for use with -120$^{\circ}$ C data that has the time-dependent gain
adjustment and CTI correction (if available) applied. There is no CTI
correction for the back-illuminated {\em ACIS} chips, {\em S1} and
{\em S3}.}, otherwise the previous tool {\em mkrmf} was used. 
For each cluster spectrum file (with or without point sources) the
respective ARFs and RMFs were generated in order to account for the
presence or absence of point sources. For each cluster spectrum file 
there are then three associated files, namely the background 
spectrum file and the two response matrices.

\smallskip\noindent 
The spectra are then analyzed with the \textit{XSPEC} package and
fitted over the energy range 0.8--7.0 keV. The photons with energy
below 0.8 keV were excluded to avoid systematic biases in the 
temperature determination due to uncertainties in the {\em ACIS} 
calibration at low energies. Ignoring energies above 7 keV has little 
effect on {\em Chandra} data due to the low effective area above that 
energy and the rapid S/N decrease of the thermal spectra. The spectra were 
fitted with an absorbed single-temperature thermal model called
\textit{wabs(mekal)} \citep{Ka92,Li95}.  The absorbing hydrogen column
was frozen at the Galactic value \citep[as determined from radio HI
maps,][]{Di90} in correspondence of the X-ray peak.  The gas
temperature, T, and the normalization, K, of the thermal component are
the only free parameters.  The best-fit temperatures were determined
freezing the redshift at the values measured by spectroscopic
observations (available in the literature) and fixing the metallicity
(Z) at 0.3 Z$_\odot$. In the spectral fitting, we take into account
the increased effective area at energies larger than 2 keV \citep[due
to a thin hydrocarbon layer, see][]{Mr03} including in the fitting
model the ``positive absorption edge'' (XSPEC model edge) described by
Eq.~1 in \cite{Vi05}.

\medskip\noindent
The spectral fit has been performed using both the Cash statistics
\citep{Ca79} and the $\chi^2$ statistics (adopting a standard binning
with a minimum of 20 photons per energy channel in the source plus
background spectrum). The Cash statistics seems to be preferable for
low-S/N spectra \citep{No89} when the number of counts available per
bin is low. However the agreement between the models obtained with the
two statistics (see Appendix~\ref{appendix:a1}) makes us confident
that both statistics are a good choice.  Throughout this paper we will
use the $\chi^2$ statistics for the only reason that, unlike the Cash
statistics, the $\chi^2$ statistics gives a measure of the absolute
goodness of the fit. We verified that all our spectra are 
well--fitted by single--temperature models: the best-fit models have a 
reduced $\chi^2 \sim$ 1 and a null-hypothesis probability above 15\% 
(except for ZW\,CL\,1454.8$+$2233, see Notes on individual clusters in
Sect.~\ref{notes}).

\smallskip\noindent
Once the models with the best-fit parameters were determined
\textit{XSPEC} has been used to calculate the cluster flux over different 
energy bands and the cluster bolometric luminosity. The errors on the 
cluster parameters are obtained in \textit{XSPEC} from the distribution of 
the values around the best-fit value of the spectral analysis.
The quoted luminosities were corrected for the effect of absorption by
the Galactic HI column density at low energies. Hereafter the
unabsorbed bolometric luminosity is called bolometric luminosity or
$L_{bol}$.

\smallskip\noindent
Since the quality of the fits is better inside the region that
maximizes the signal to noise ratio, the temperatures estimated within
$\rm R_{spec}$, i.e. $\rm T_{R_{spec}}$, are considered more
representative of the actual average temperature of the gas.  However
the two temperatures are consistent within the errors with no
significant systematic differences between them.  For the luminosities
we used instead the $\rm R_{ext}$ radius in order to take into account
the faint brightness tails at the cluster periphery as well as the
point sources in those regions. The choice of the different radii
($\rm R_{spec}$ for the determination of temperature and $\rm R_{ext}$
for the determination of luminosity) is justified in
Appendix~\ref{appendix:a2} where a more detailed description of the
analysis of the best-fit cluster parameters is given.

\medskip\noindent
The cluster parameters are listed in Table~\ref{tab2}. 
The columns contain the following information:
\begin{itemize}
\item [--] Column 1: The first line gives the cluster name, the second 
               and third lines indicate how parameters are derived, if for the 
               cluster alone or for the cluster plus point sources. An asterisk 
               close to the cluster name indicates that the cluster has been 
               classified as a possible cooling core by \cite{Vi02}
\item [--] Column 2-3: Core radius in kpc and $\beta$
\item [--] Column 4: Radius (spec) which maximizes the S/N, in arcsec and kpc 
\item [--] Column 5: Radius (ext) where the cluster X-ray radial profile 
               becomes flat, in arcsec and kpc
\item [--] Column 6: Number of point sources detected (and removed) within
                   $\rm R_{spec}$ and $\rm R_{ext}$, respectively. These 
                   point sources were detected either in the soft or in the 
                   hard energy band (see BR07 for details). The column lists 
                   also faint sources which become significant 
                   ($\rm S/N > 3$) in the full energy band (0.5--7.0 keV).
\item [--] Column 7: Temperature estimated within $\rm R_{spec}$ in keV
\item [--] Column 8: Temperature estimated within $\rm R_{ext}$ in keV
\item [--] Column 9-10: Observed soft and hard flux estimated within 
               $\rm R_{ext}$ in units of 10$^{-13}$ erg cm$^{-2}$ s$^{-1}$
\item [--] Column 11: Unabsorbed X-ray bolometric luminosity estimated within  
               $\rm R_{ext}$ in units of  10$^{44}$ erg s$^{-1}$
\end{itemize}
\tabcolsep 0.1cm
\begin{longtable}{lcccccrrrrr}
\caption{X-ray Cluster Parameters}
\
\\
\hline
Cluster & $\rm R_{c}$  & $\beta$ &$\rm R_{spec}$  & $\rm R_{ext}$ & $\rm N_{sources} $  &T$_{\rm R_{spec}}$& T$_{\rm R_{ext}}$ & S$_{0.5-2.0}$ & S$_{2.0-10.0}$  & L$_{bol}$\\
     & (kpc) & & (\arcsec)~(kpc) & (\arcsec)~(kpc) & ($\rm R_{spec}$)~($\rm R_{ext}$) & (keV)   & (keV) & (10$^{-13}$ cgs) & (10$^{-13}$ cgs) & (10$^{44}$ cgs)\\ 
(1) & (2) & (3) & (4) & (5) & (6) & (7) & (8) & (9) & (10) & (11) \\
\hline
\hline
\endhead
Abell\,2125 &182$\pm$12 & 0.54$\pm$0.02&   153~~591     &  241~~931   & 13~~~~26 &                    &                       &                               &                        &                         \\ 
cl    & &       &                &  &           & $3.4_{-0.2}^{+0.2}$ & $ 3.5_{-0.2}^{+0.2}$  &$ 4.06_{-0.14}^{+0.14}$        &$ 4.13 _{-0.22}^{+0.21}$&$  2.20_{-0.07}^{+0.06}$ \\  
cl+ps & &       &                &  &           & $3.5_{-0.1}^{+0.1}$ & $ 3.7_{-0.2}^{+0.3}$  & $ 4.79_{-0.14}^{+0.14}$       &$5.11_{-0.23}^{+0.18}$  & $2.64_{-0.07}^{+0.07}$  \\ 
\hline
ZW\,CL\,1454.8$+$2233& & & 128~~512       &  200~~800   &   4~~~~~7  &                &                       &                        &                        &                         \\ 
cl  & &         &                &  &           & $4.4_{-0.1}^{+0.1}$ & $ 4.5_{-0.1}^{+0.1}$  &$32.05_{-0.18}^{+0.18}$ &$ 39.71_{-0.47}^{+0.44}$&$ 20.83_{-0.10}^{+0.10}$ \\  
cl+ps & &       &                &  &           & $4.4_{-0.1}^{+0.1}$ & $ 4.5_{-0.1}^{+0.1}$  &$32.34_{-0.18}^{+0.17}$& $40.08_{-0.41}^{+0.42}$ & $ 21.02_{-0.11}^{+0.10}$ \\ 
\hline
MS\,1008.1$-$1224&165$\pm$9  & 0.64$\pm$0.02&  128~~572      &  172~~771   &  6~~~~10 &                   &                       &                        &                        &                         \\ 
cl  & &         &                &  &           & $6.0_{-0.3}^{+0.4}$ & $ 6.1_{-0.4}^{+0.4}$ &$ 8.76_{-0.21}^{+0.19}$ &$ 15.67_{-0.82}^{+0.60}$&$ 10.28_{-0.34}^{+0.29}$ \\  
cl+ps & &       &                &  &           & $6.2_{-0.4}^{+0.4}$ & $ 6.3_{-0.4}^{+0.4}$  & $ 8.97_{-0.21}^{+0.20}$& $16.32_{-0.71}^{+0.63}$& $ 10.65_{-0.33}^{+0.30}$ \\ 
\hline
ZW\,CL\,0024.0$+$1652$^{*}$&128$\pm$10 & 0.67$\pm$0.03&  69~~367       &  118~~628   &  2~~~~~7 &                   &                       &                        &                        &                         \\ 
cl  & &         &               &  &           & $4.4_{-0.4}^{+0.5}$ & $ 4.4_{-0.5}^{+0.7}$  &$ 2.25_{-0.14}^{+0.14}$ &$ 2.67 _{-0.28}^{+0.25}$&$  3.90_{-0.22}^{+0.22}$ \\    
cl+ps & &           &           &  &           & $4.8_{-0.6}^{+0.6}$ & $ 5.0_{-0.7}^{+0.7}$  & $ 2.50_{-0.15}^{+0.12}$& $ 3.35_{-0.40}^{+0.33}$& $ 4.55_{-0.29}^{+0.26}$\\
\hline	     
MS\,1621.5$+$2640 &227$\pm$17 & 0.65$\pm$0.03&  118~~659      & 148~~823    &  3~~~~~5 &                   &                       &                        &                        &                         \\ 
cl  & &           &    &         &              & $7.5_{-0.7}^{+1.1}$ & $ 7.5_{-0.8}^{+1.3}$  &$ 4.31_{-0.23}^{+0.21}$ &$ 7.78 _{-0.59}^{+0.41}$&$ 11.07_{-0.56}^{+0.54}$ \\      
cl+ps & &         &            & &              & $7.8_{-0.8}^{+1.1}$ & $ 8.0_{-1.0}^{+1.1}$  & $ 4.79_{-0.20}^{+0.18}$& $ 8.99_{-0.77}^{+0.66}$&$ 12.69_{-0.82}^{+0.79}$ \\ 
\hline
RXJ\,1701.3$+$6414$^{*}$& 15$\pm$2  & 0.41$\pm$0.01&  79~~455          &  108~~626         & 2~~~~~3                    &                       &                        &                        &                 \\ 
cl  & &           &	  &          &              & $4.5_{-0.3}^{+0.4}$ & $ 5.0_{-0.5}^{+0.6}$  &$ 2.61_{-0.13}^{+0.14}$ &$ 3.26 _{-0.31}^{+0.24}$&$  6.27_{-0.33}^{+0.28}$ \\    
cl+ps & &            &            &  &               & $4.8_{-0.4}^{+0.5}$ & $ 5.5_{-0.6}^{+0.6}$  & $ 2.62_{-0.15}^{+0.15}$& $ 3.55_{-0.30}^{+0.26}$& $ 6.54_{-0.37}^{+0.28}$ \\ 
\hline
CL\,1641$+$4001&151$\pm$18 & 0.77$\pm$0.06&  54~~317          &  89~~519         &   3~~~~~6 &                   &                       &                       &                        &                         \\ 
cl & &            &	  &          &              & $5.1_{-0.7}^{+0.8}$ & $ 5.5_{-0.9}^{+1.1}$   &$ 1.05_{-0.11}^{+0.10}$  &$ 1.36 _{-0.19}^{+0.14}$&$  2.65_{-0.22}^{+0.22}$ \\    
cl+ps  & &           &            &  &          & $4.9_{-0.6}^{+0.6}$ & $ 5.7_{-0.7}^{+0.9}$  & $ 1.57_{-0.13}^{+0.12}$& $ 2.09_{-0.19}^{+0.15}$  & $  4.02_{-0.26}^{+0.22}$\\ 
\hline
V\,1524.6$+$0957&302$\pm$27 & 0.80$\pm$0.05&   79~~488          &   148~~916        & 3~~~~~10 &                    &                       &                        &                        &                         \\ 
cl    & &       &	 &         &                & $5.0_{-0.5}^{+0.6}$ & $ 5.6_{-0.8}^{+1.1}$ &$ 2.00_{-0.20}^{+0.18}$ &$ 2.72 _{-0.33}^{+0.22}$&$  6.85_{-0.52}^{+0.44}$ \\	   
cl+ps & &       &            &     &            & $5.2_{-0.6}^{+0.6}$ & $ 6.4_{-0.9}^{+1.1}$ & $ 2.39_{-0.17}^{+0.16}$& $3.57_{-0.44}^{+0.32}$& $  8.58_{-0.64}^{+0.50}$\\ 
\hline
MS\,0451.6$-$0305&270$\pm$8  & 0.90$\pm$0.02& 89~~562            &  148~~937         &   3~~~~~6 &                &                       &                        &                        &                         \\ 
cl    & &       &	 &        &                 & $9.4_{-0.5}^{+0.7}$ & $9.6_{-0.7}^{+1.0}$ & $9.81_{-0.20}^{+0.20}$ &$ 20.84_{-1.41}^{+1.12}$& $ 50.43_{-2.67}^{+2.56}$ \\    
cl+ps  & &      &            &    &             & $9.3_{-0.5}^{+0.6}$ & $ 9.8_{-0.8}^{+1.0}$& $ 9.96_{-0.24}^{+0.19}$& $21.41_{-1.44}^{+0.99}$& $ 51.77_{-2.44}^{+2.44}$ \\ 
\hline
V\,1121$+$2327&437$\pm$58 & 1.19$\pm$0.18     & 67~~434           & 128~~829          &  4~~~~~8 &                   &                        &                        &                         \\ 
cl  & &         &	  &       &                 & $4.5_{-0.4}^{+0.5}$ & $ 5.5_{-0.9}^{+1.1}$ &$ 1.38_{-0.13}^{+0.11}$ &$ 1.72 _{-0.23}^{+0.19}$&$  5.44_{-0.43}^{+0.37}$ \\    
cl+ps & &       &            &    &               & $4.5_{-0.4}^{+0.5}$ & $ 5.9_{-0.8}^{+1.2}$ & $ 1.58_{-0.14}^{+0.13}$& $ 2.08_{-0.26}^{+0.15}$& $  6.38_{-0.46}^{+0.35}$ \\ 
\hline
MS\,2053.7$-$0449$^{*}$&115$\pm$12 & 0.64$\pm$0.03&  57~~373          &  118~~779         &  1~~~~~3                   &                       &                        &                        &                         \\ 
cl   & &        &	  &     &                   & $4.3_{-0.4}^{+0.5}$ & $ 5.1_{-1.0}^{+1.4}$ &$ 1.26_{-0.15}^{+0.12}$ &$ 1.51 _{-0.33}^{+0.19}$&$  5.69_{-0.61}^{+0.46}$ \\     
cl+ps  & &      &            &  &               & $4.5_{-0.4}^{+0.6}$ & $ 5.2_{-1.0}^{+1.4}$   & $ 1.28_{-0.14}^{+0.13}$& $ 1.56_{-0.36}^{+0.20}$& $  5.81_{-0.60}^{+0.43}$ \\ 
\hline
V\,1221$+$4918&272$\pm$20 & 0.76$\pm$0.04    & 79~~562            &  143~~1020         &  3~~~~~8 &                   &                       &                        &                        &                         \\ 
cl   & &        &	 &      &                   & $7.0_{-0.7}^{+0.8}$ & $ 6.4_{-1.0}^{+0.9}$ &$ 1.90_{-0.11}^{+0.10}$ &$ 2.49 _{-0.27}^{+0.25}$&$ 13.12_{-0.75}^{+0.73}$ \\     
cl+ps & &       &            &  &               & $7.2_{-0.3}^{+0.8}$ & $ 7.2_{-0.8}^{+1.1}$   & $ 2.19_{-0.12}^{+0.11}$& $ 3.16_{-0.26}^{+0.20}$& $ 15.87_{-0.78}^{+0.69}$ \\ 
\hline
MS\,1137.5$+$6625&116$\pm$6  & 0.71$\pm$0.02   & 59~~440           & 103~~770          &  1~~~~~6 &                   &                       &                        &                        &                         \\ 
cl   & &        &	  &     &                   & $6.2_{-0.4}^{+0.6}$ & $ 5.7_{-0.5}^{+0.6}$ &$ 1.61_{-0.08}^{+0.09}$ &$ 1.82 _{-0.12}^{+0.10}$&$ 13.62_{-0.54}^{+0.48}$ \\     
cl+ps & &       &            &  &               & $6.3_{-0.5}^{+0.6}$ & $ 6.0_{-0.4}^{+0.6}$   & $ 1.99_{-0.08}^{+0.07}$& $ 2.37_{-0.18}^{+0.15}$& $ 17.17_{-0.74}^{+0.58}$ \\ 
\hline
RDCSJ\,1317$+$2911$^{*}$& 61$\pm$16 & 0.52$\pm$0.04&  30~~222            &   69~~518           &   1~~~5 &                  &                       &                        &                        &                         \\ 
cl   & &        &	 &       &                  & $3.7_{-0.8}^{+1.2}$ & $ 2.4_{-0.6}^{+0.9}$ &$ 0.17_{-0.05}^{+0.04}$ &$ 0.06 _{-0.03}^{+0.02}$&$  1.28_{-0.37}^{+0.36}$ \\     
cl+ps  & &      &            &   &              & $5.8_{-1.6}^{+2.9}$ & $ 4.1_{-1.3}^{+1.8}$  
& $ 0.20_{-0.04}^{+0.04}$& $ 0.15_{-0.06}^{+0.02}$ & $  1.62_{-0.31}^{+0.23}$ \\ 
\hline
RDCSJ\,1350$+$6007&261$\pm$43 & 0.70$\pm$0.07& 64~~481           & 128~~962         &  3~~~~~9 &                   &                       &                        &                        &                         \\ 
cl    & &       &	 &       &                  & $4.1_{-0.6}^{+0.8}$ & $ 3.7_{-0.7}^{+1.2}$ &$ 0.77_{-0.13}^{+0.12}$ &$ 0.51 _{-0.12}^{+0.06}$&$  6.19_{-0.88}^{+0.76}$ \\
cl+ps & &       &            &   &              & $4.6_{-0.7}^{+0.9}$ & $ 4.6_{-0.8}^{+1.3}$  & $ 0.96_{-0.14}^{+0.12}$& $ 0.84_{-0.15}^{+0.09}$& $  8.14_{-0.93}^{+0.73}$  \\ 
\hline
RXJ\,1716.4$+$6708&119$\pm$11 & 0.66$\pm$0.03 & 59~~446           & 108~~817          &  3~~~~~9 &                   &                       &                        &                        &                         \\ 
cl    & &       &	  &      &                  & $6.5_{-0.8}^{+0.9}$ & $ 6.0_{-0.7}^{+1.1}$ &$ 1.31_{-0.11}^{+0.09}$ &$ 1.62 _{-0.27}^{+0.22}$&$ 13.15_{-1.16}^{+0.93}$ \\     
cl+ps & &       &            &   &              & $7.8_{-0.9}^{+1.2}$ & $ 8.1_{-1.2}^{+1.7}$   & $ 1.47_{-0.11}^{+0.10}$& $ 2.35_{-0.36}^{+0.25}$& $ 16.76_{-1.52}^{+1.22}$ \\ 
\hline
MS\,1054.4$-$0321&520$\pm$32 & 1.38$\pm$0.11    &   84~~636         &    128~~972       &  2~~~~~6                   &                       &                        &                        &                         \\ 
cl   & &        &	  &      &                  & $8.3_{-0.7}^{+0.7}$ & $ 7.8_{-0.9}^{+1.0}$ &$ 2.96_{-0.12}^{+0.10}$ &$ 4.59 _{-0.37}^{+0.37}$&$ 34.83_{-1.82}^{+1.66}$ \\     
cl+ps & &       &            &   &              & $8.9_{-0.7}^{+0.7}$ & $ 8.6_{-0.9}^{+0.9}$   & $ 3.13_{-0.10}^{+0.10}$& $ 5.17_{-0.38}^{+0.34}$& $ 38.26_{-1.75}^{+1.56}$ \\ 
\hline
WARPJ\,1415.1$+$3612& 68$\pm$7  & 0.60$\pm$0.02 &  39~~316          & 79~~632           & 2~~~~~4  &                   &                       &                        &                        &                         \\ 
cl   & &        &	  &         &               & $6.2_{-0.7}^{+0.8}$ & $ 6.3_{-0.9}^{+1.0}$ &$ 0.67_{-0.06}^{+0.05}$ &$ 0.75 _{-0.12}^{+0.07}$&$ 11.88_{-0.91}^{+0.71}$ \\     
cl+ps & & 	    &            &  &               & $7.0_{-0.8}^{+0.9}$ & $ 7.1_{-1.0}^{+1.3}$  & $ 0.79_{-0.07}^{+0.06}$& $ 0.97_{-0.11}^{+0.06}$& $ 13.10_{-0.96}^{+0.83}$ \\ 
\hline
\label{tab2}
\end{longtable}
\section{Comparison with other authors}
\label{comparison}

Two recent works, one by \cite{Vi02} (from now on VI02) and one by
\cite{Et04} (from now on ET04), have 13 and 12 clusters respectively
in common with us. It is thus very instructive to compare our
estimates of the gas temperatures and bolometric luminosities with
their results in order to check for the presence of any systematic
bias due to the different approaches adopted. Of particular importance
is the radius used to measure the properties of the clusters.

\subsection{Temperatures}

\begin{figure}
\begin{center}
\resizebox{\textwidth}{!}
{\includegraphics{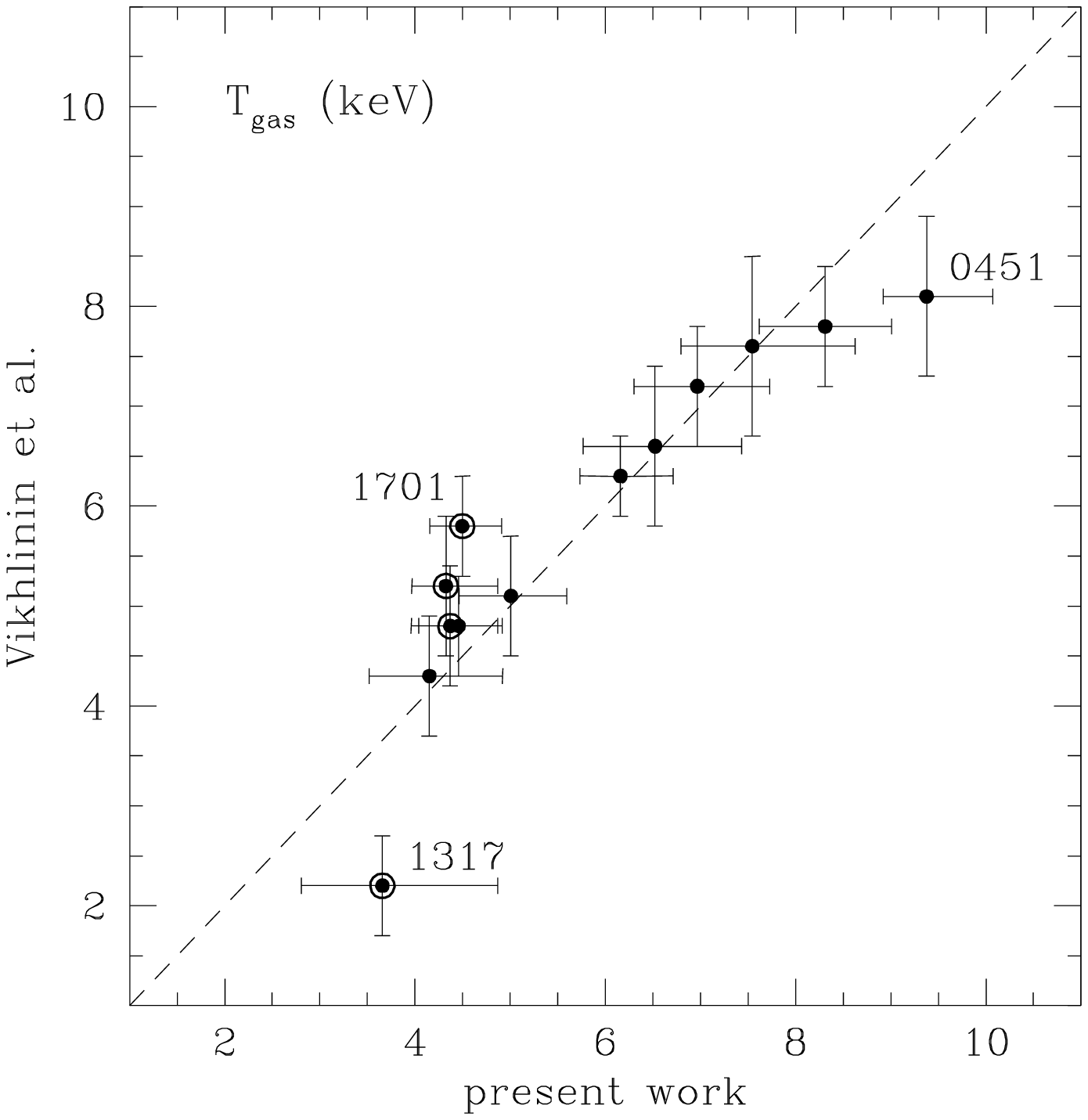}\includegraphics{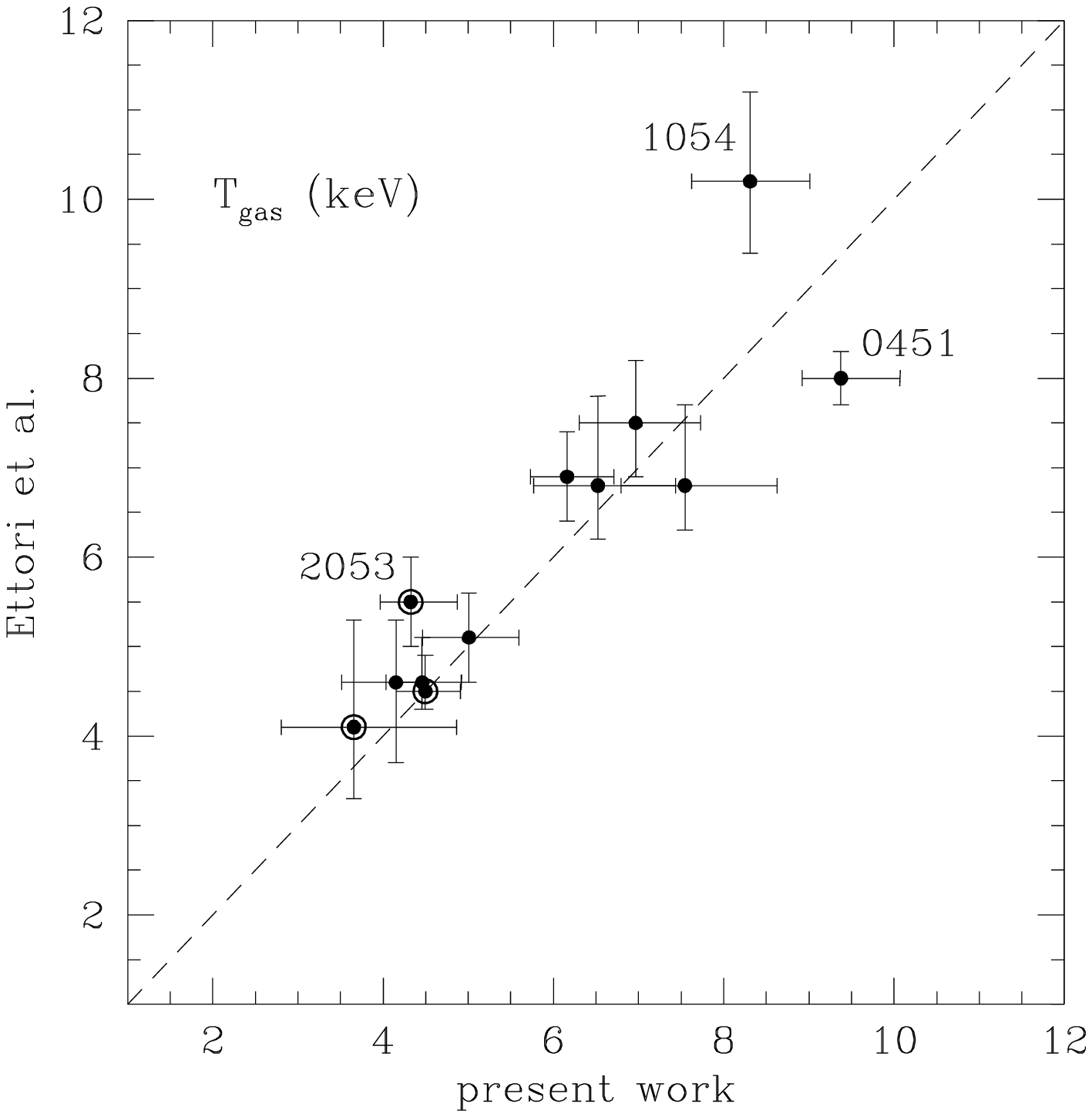}}
\end{center}
\caption{Comparison between the estimates of the gas temperature 
by  VI02 (\textit{left panel}) and our estimate for the 13 clusters 
in common. To the \textit{right panel} the same comparison is shown
for the 12 clusters in common  with ET04.  The circled points indicate 
clusters assumed by VI02 to be cooling core clusters.  Like us, ET04 
did not  excise any excess surface brightness central region due to the 
presence of ``cooling flows''.  In both panels the dashed line is equality 
between the two works.}
\label{fig1}
\end{figure}

A comparison of our temperatures with those of VI02 is shown in
Fig.~\ref {fig1} (panel to the left). Differently from us, VI02
excluded the central $100 ~h^{-1}_{50}$ kpc region in the cooling core
clusters.  The agreement is very good: a mean ratio of 1.02$\pm$0.04
between VI02 temperatures and ours has been found. In three out of
four clusters assumed by VI02 to be possible cooling core (indicated
by an asterisk in Table~\ref{tab2}) they report higher temperatures.
The largest difference, significant at 2.9$\sigma$, is found for
RXJ\,1701$+$6414.  For the fourth cluster, RDCSJ\,1317$+$2911, we find
instead a higher temperature, $k \rm T=3.7^{+1.2}_{-0.8}$ keV against
the value found by VI02 of $k \rm T=2.2^{+0.5}_{-0.5}$ keV. The
difference is significant at the 2.2$\sigma$ level. For
MS\,0451.6$-$0305 the temperature difference, significant at 2$\sigma$
confidence level, can be explained in terms of the new calibrations
that we have applied.
\noindent

\smallskip\noindent
The comparison with ET04 (their values vs. ours) shows that their
temperatures are on average a factor of 1.06$\pm$0.03 higher than ours
(see panel to the right in Fig.~\ref{fig1}). The largest absolute
values of the differences (significant at $>2\sigma $) are found for
MS\,0451.6$-$0305 ($3.6\sigma$), MS\,1054$-$0321 ($2.5\sigma$) and
MS\,2053.7$-$0449 (2.2$\sigma$).  All these discrepancies are
discussed in the notes on individual clusters (Sect.~\ref{notes}).

\subsection{Luminosities}

It has to be noted that both VI02 and ET04 derived cluster
luminosities within areas often quite different from the areas we
used. Many authors discuss the importance of the choice of the radius
within which cluster properties are measured, especially when
comparing integrated cluster properties with theoretical predictions
or simulations.  A common choice is to use as radius a fixed linear 
size, which has the obvious benefits in terms of simplicity.
VI02 used a fixed radius of $2 ~h^{-1}_{50}$ Mpc, corresponding to
$1.4 ~h^{-1}_{70}$ Mpc in our cosmology. Since VI02 excluded the
central $100 ~h^{-1}_{50}$ kpc regions in the cooling core clusters,
they accounted for the missed flux by multiplying by a factor
1.06 typical of a $\beta$-model cluster.

\begin{figure}
\begin{center}
\resizebox{\textwidth}{!}{\includegraphics{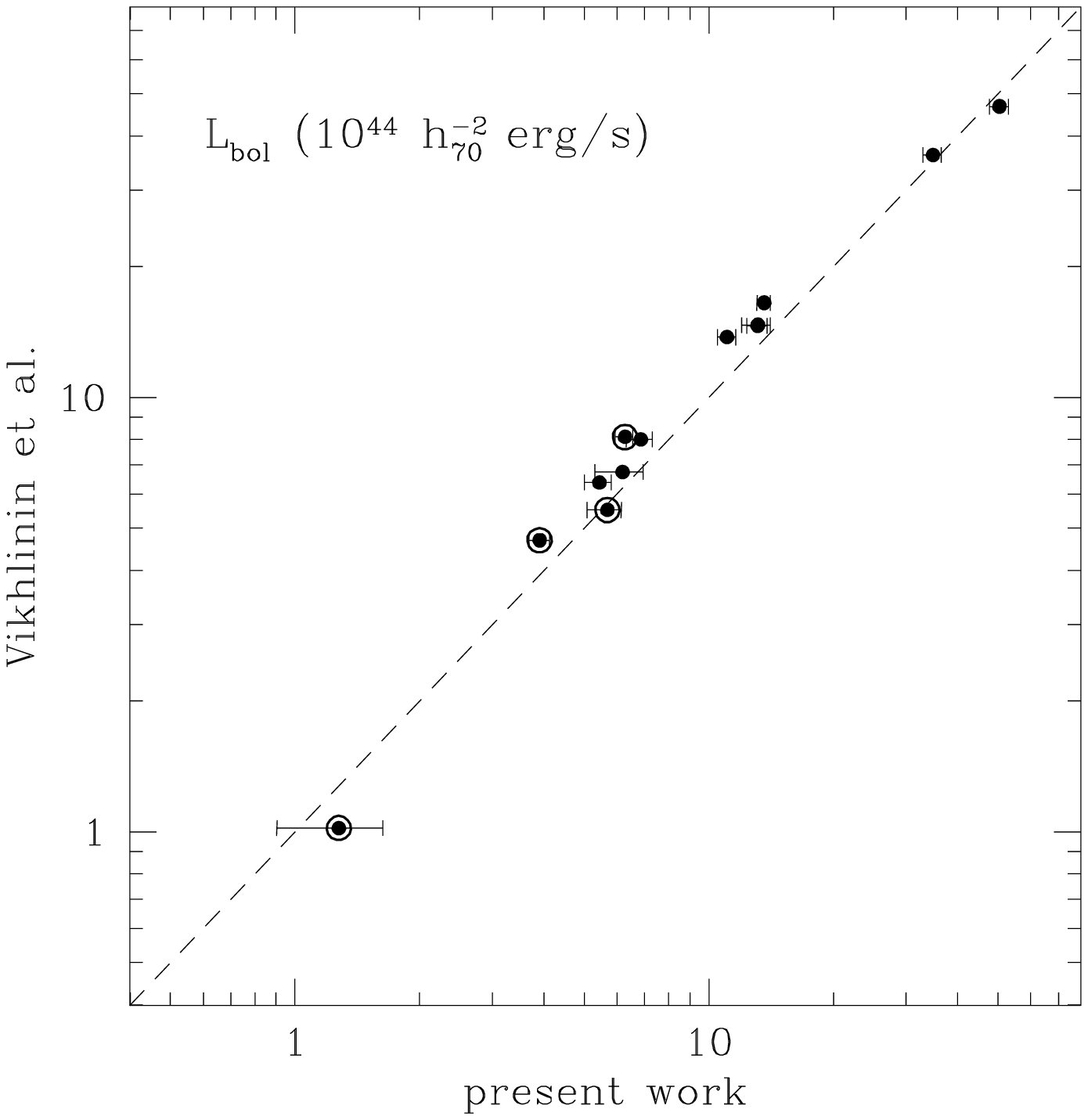}
\includegraphics{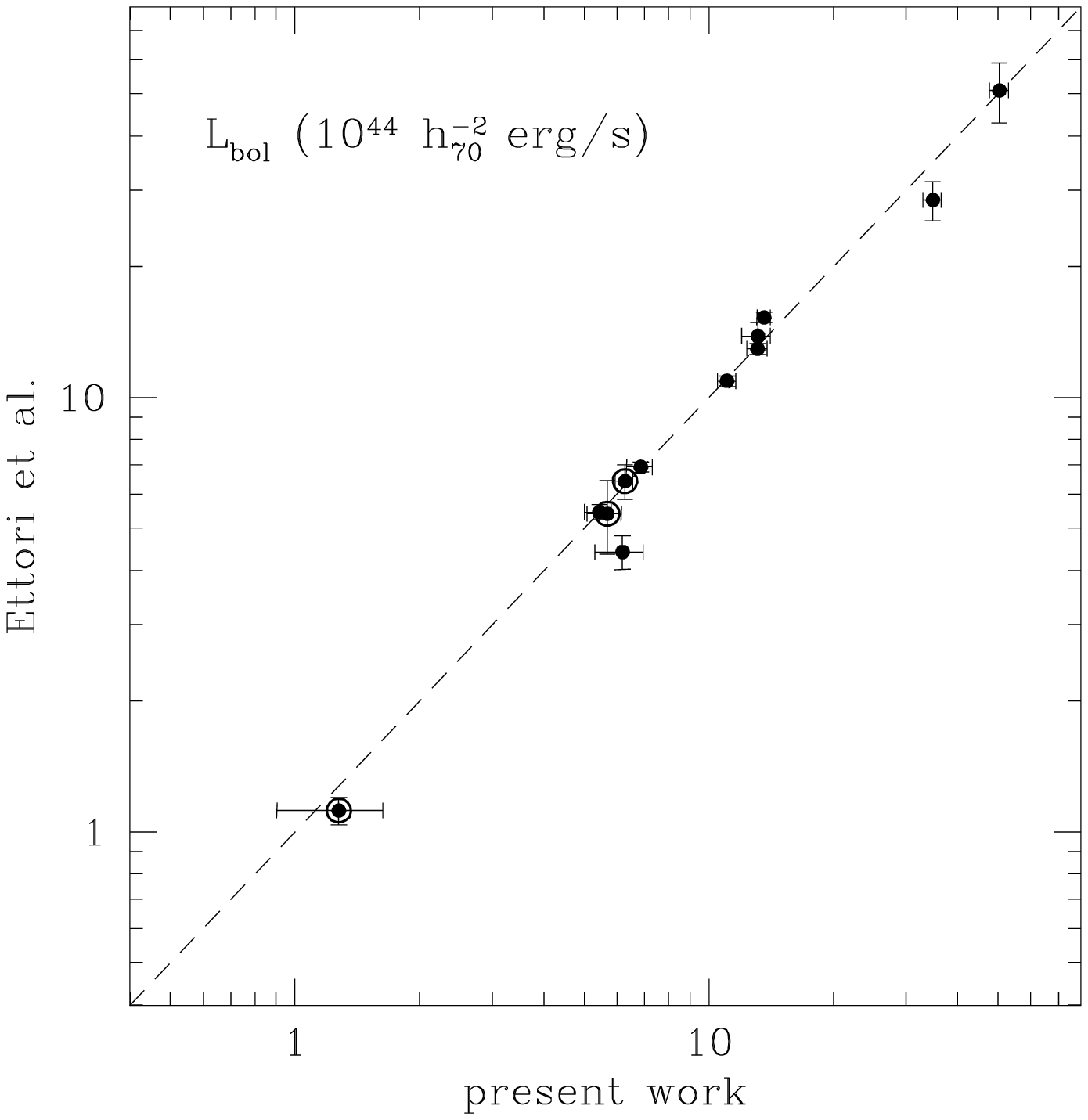}}
\end{center}
\caption {Comparison between the luminosities estimated by 
VI02 and by us for the 13 clusters in common (\textit{left panel})
and by ET04 and by us for the 12 clusters in common (\textit{right 
panel}). The circled points indicate the clusters assumed by VI02 to be
cooling core clusters. In both panels the dashed line is equality 
between the two works.}
\label{fig2}
\end{figure}

\smallskip\noindent 
Other authors define a ``more physical" radius
which requires the knowledge of the cluster mass profile so that the
mean enclosed density is a fixed factor above the critical density of
the Universe. This approach is followed by ET04 who define $\rm
R_{500}$, which corresponds to an overdensity $\Delta_{z} = 500 \times
\Delta_{v}(z)/(18\pi^2)$ with respect to the critical density of the
Universe at redshift z.  Their luminosities were computed by
extrapolating to $\rm R_{500}$ the values measured within $\rm
R_{spec}$ by means of an isothermal $\beta$-profile. Like us, ET04 did
not excise any excess surface brightness central region due to the
presence of ``cooling flows''.  Differently from these authors we used
an ``observed'' radius which indicates the region where the emission
is detected ($\rm R_{ext}$, see Sect.~\ref{radius}). The VI02 radii
differ from our radii by a factor ranging from 1.04 up to 2.7. Also
the ET04 radii are larger than our radii by a factor 
ranging from 1.03 up to 1.55, except for RDCSJ\,1350$+$6007 that has
a ET04 radius about 30\% less than ours.

\medskip\noindent
The comparison between the VI02 luminosities and ours gives a mean
ratio of 1.10$\pm$0.04 (see left panel of Fig.~\ref{fig2}).  
The fact that the VI02 luminosities are on average 10\% higher than 
our estimates may be explained by the larger regions (radius of 2 
$h_{50}^{-1}$ Mpc) used by them. The mean ratio between the ET04 
luminosities and ours is 0.96$\pm$0.03 (see right panel of Fig.~\ref{fig2}).

\begin{figure}
\centering
\includegraphics[width=12cm,height=12cm,bb=0 30 574 604]{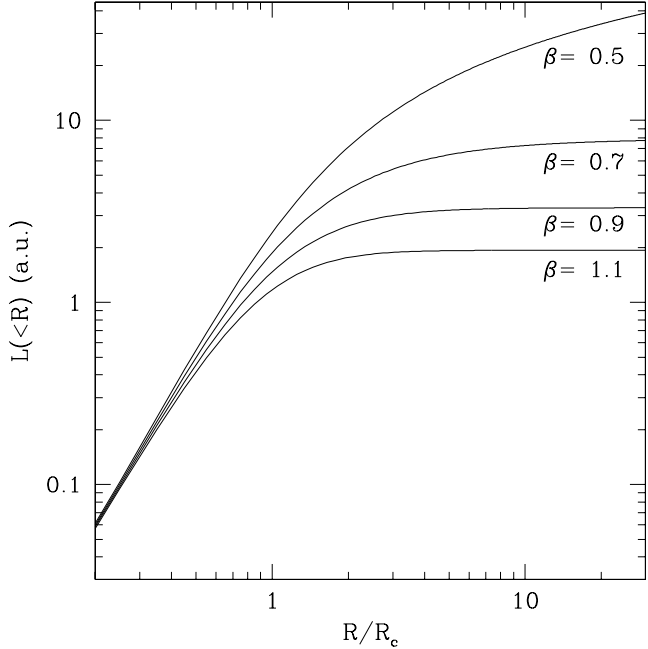}
\vspace{-3cm}
\caption{The luminosity computed within a radius R as a function of
the ratio between the radius R and the core radius $\rm R_{c}$ for different 
$\beta$ values.}
\label{fig3}
\end{figure}

\smallskip\noindent
To check if any systematic bias is introduced by our choice of the 
radius, we extrapolate our luminosities to the radii adopted by the
other authors. The correction to be applied to our data was
obtained assuming that the gas density profile is described by 
the $\beta$-model (see Eq.~\ref{cl:eqn2}) defined by our estimates for 
the core radius, $\rm R_{c}$, and for $\beta$ (see Sect.~\ref{spatial}). 
It has to be noted that these corrections do not
depend on the ratio between the different integration radii 
only but depend strongly on $\beta$ and $R_{c}$. This can be clearly seen 
in Fig.~\ref{fig3} which shows the luminosity computed within a given 
radius R as a function of R/R$_{c}$, for different values of $\beta$.

\begin{figure}
\begin{center}
\resizebox{\textwidth}{!}
{\includegraphics{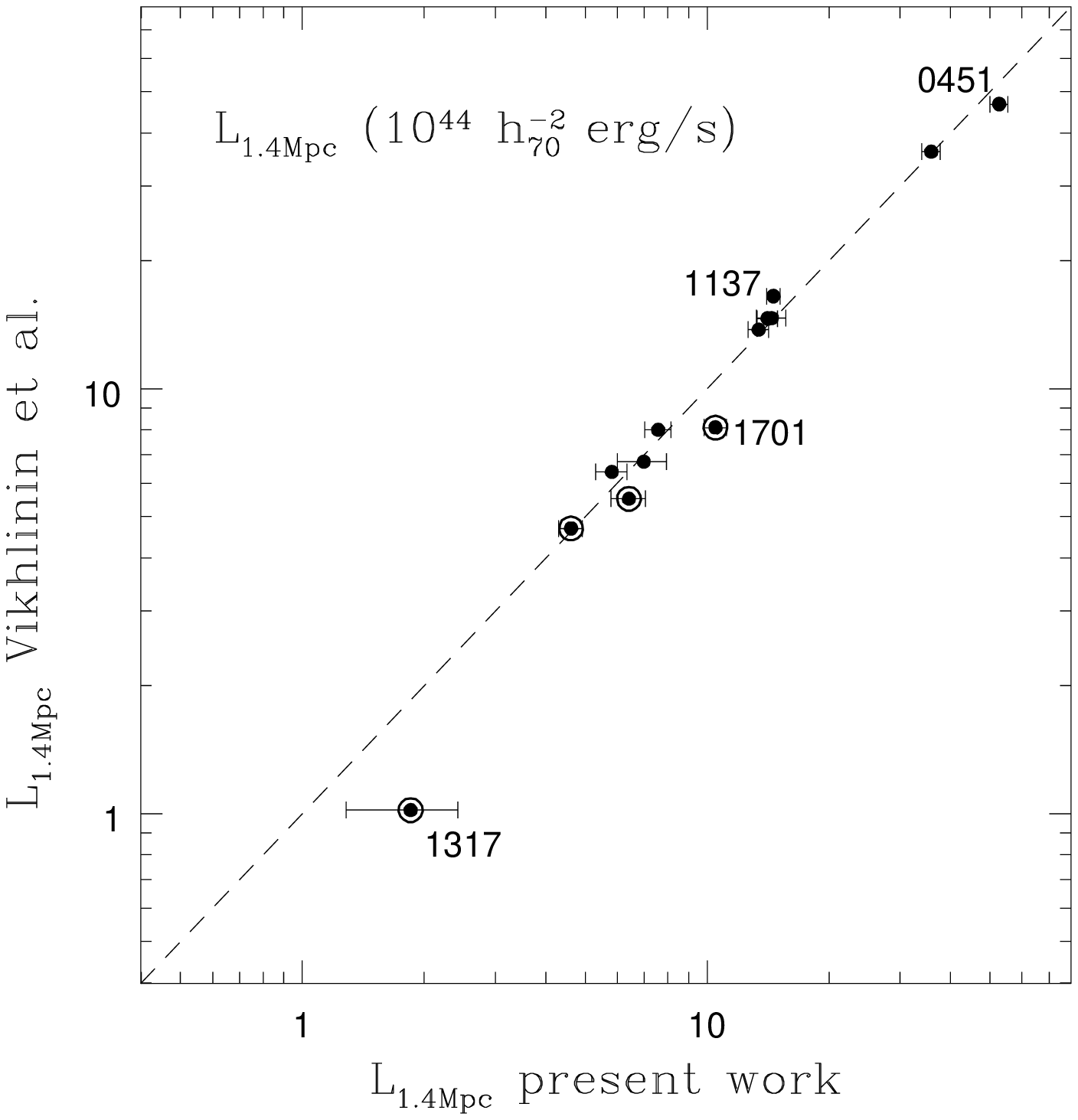}
\includegraphics{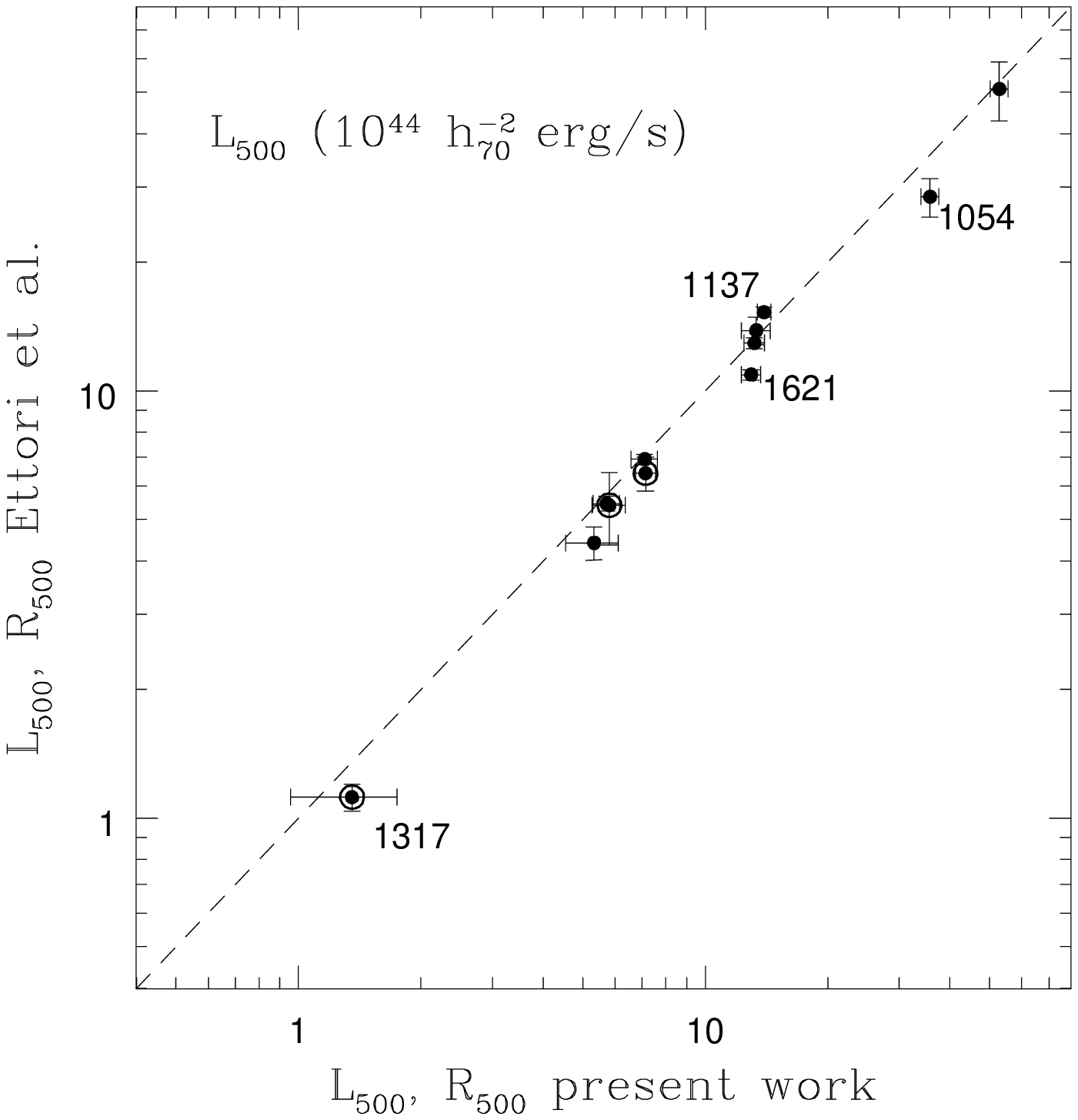}}
\end{center}
\caption{(\textit{Left panel}) Comparison between the luminosities
estimated by VI02 and our luminosities extrapolated to 1.4
h$^{-1}_{70}$ Mpc for the 13 clusters in common. The circled points
indicate the clusters assumed by VI02 to be cooling core
clusters. (\textit{Right panel}) Comparison between the luminosities
estimated by ET04 and our luminosities extrapolated to $\rm R_{500}$
for the clusters in common. In both panels the uncertainties on the
extrapolated luminosities are combined errors which take into account
both the errors obtained in the spectral analysis by XSPEC and the
uncertainties on the $\beta$-model for the gas density profile. In both 
panels the dashed line is equality between the two works.}
\label{fig4}
\end{figure}

\smallskip\noindent 
The comparison between VI02 luminosities and our luminosities 
extrapolated to 1.4 h$^{-1}_{70}$ Mpc gives a very good agreement
(see left panel of Fig.~\ref{fig4}). Excluding the most 
discrepant cluster in the plot, RDCSJ\,1317$+$2911 (which has although
very large errorbar values) the mean ratio between the VI02 luminosities 
and ours is of 0.99$\pm$0.03. Note that the corrections to be applied 
to our luminosities for such an extrapolation are 
not negligible.  They  range from $\sim$ 0.9 to $\sim$ 1.25 with 
a median  value of 1.10 (mean 1.13). Similar results 
were obtained  using the VI02 estimates of $R_{c}$ and $\beta$,
instead of ours, to extrapolate to the VI02 radius.
The clusters with the largest absolute values ($>2\sigma $)
of the difference between our and VI02 luminosities are indicated in the
figure: the possible cooling core cluster RXJ\,1701.3$+$6414
($4.0\sigma$), MS\,1137.5+6625 ($3.2\sigma$) and MS\,0451.6-0305 ($2.3\sigma$).

%

%
\smallskip\noindent 
In the comparison with ET04 (see right panel in Fig.~\ref{fig4}) we
extrapolated our luminosities to $\rm R_{500}$. This radius was
calculated assuming our estimates for the cluster temperature, for the
core radius and $\beta$ (see Paper II of Branchesi et al. 2007 for
more details). If one compares the right panel of Fig.~\ref{fig4} with
the right panel of Fig.~\ref{fig2}, it can be noted that the
extrapolation to $\rm R_{500}$ does not improve the comparison with
ET04.  In fact the mean ratio between the ET04 luminosities and ours
changes from 0.96$\pm$0.03 to 0.93$\pm$0.02. A similar result is
obtained if we extrapolate our luminosities using their
estimates for $\rm R_{500}$, $\rm R_{c}$ and $\beta$. In the right
panel in Fig.~\ref{fig4} cluster luminosities which deviate  
more than $>2\sigma$ from the ET04 estimates are indicated 
with the cluster name. They are MS\,1621.5+2640 ($4.0\sigma$), 
MS\,1054.4-0321 ($3.0\sigma$), MS\,1137.5+6625 ($2.7\sigma$).

\smallskip\noindent
In summary the comparison with VI02 suggests that the correction for
the different radii is important if we want to be consistent with
their luminosities. Their radius is in fact quite larger than ours,
but after corrections are implemented the agreement is very good.  The
comparison with ET04 instead suggests that the correction for the
different radii are smaller than the measurement uncertainties and
hence could be neglected. The agreement between our and ET04
measurements is good, although there is a very weak systematic offset
in the sense that ET04 luminosities tend to be lower than our
estimates.

\noindent
In paper II of Branchesi et al. 2007 we describe in detail why it is
important to extrapolate all luminosities to an homogeneous
radius (e.g., $\rm R_{500}$) when the observed L$_{bol}$--T relations 
are compared with the self-similarity evolution predictions.

\section{Notes on individual clusters}
\label{notes}
\subsection {MS\,0451.6$-$0305}

MS\,0451.6$-$0305 is the most luminous cluster in the EMSS sample
\citep{Gio90}. We found a best-fit temperature within $\rm R_{spec}$
of $k\rm T=9.4^{+0.7}_{-0.5} ~keV$, which disagrees with the estimates
of $k\rm T=8.1^{+0.8}_{-0.8} ~keV$ and of $k\rm T=8.0^{+0.3}_{-0.3}
~keV$ found by VI02 and ET04, respectively (see Fig.~\ref{fig1}).

\smallskip\noindent
In a more recent article \cite{Do03} analyzed the same
\textit{Chandra} data. They discuss in detail the results of applying
a soft-energy, time-dependent correction to the {\em ACIS-S}, which
however they consider uncertain. Thus without applying the correction
the authors find that MS\,0451.6-0305 is consistent with an isothermal
cluster with $k$T ranging from 10 keV to 10.6 keV ($\pm$ 1.6 keV at
the 90\% confidence level), and with intracluster Fe abundance range
between 0.32 and 0.40 ($\pm$ 0.13 solar at the 90\% confidence level).
Including the correction in their analysis, they may explain the 
discrepancy between their best-fit temperature and the temperature 
obtained by VI02. However they find that to be acceptable the fit 
requires a second component that could be either a
cooler thermal component or a steep power-law component. 

\smallskip\noindent
Our data have been analyzed applying the time-dependent correction as
suggested by the CXC ({\em Chandra} X-ray Center).  Our correction is
more accurate since the CALDB used by us is more recent than the one
used by \cite{Do03}. The more recent calibration adopted can 
explain the temperature discrepancies with respect to VI02 and ET04
(see Fig.~\ref{fig1}).  Our estimate of $k\rm T=9.4^{+0.7}_{-0.5}
~keV$ is consistent with the {\em ASCA} measure of $k\rm
T=10.2^{+1.5}_{-1.3} ~keV$ \citep{Mu97}.  We detected six very faint
sources ($\rm S_{0.5-1~0keV} < 10^{-14} \funits$) within the more
extended aperture radius $\rm R_{ext} = 100''$ (see BR07). The cluster
temperature does not change much if one includes the point sources
($k\rm T=9.6^{+1.0}_{-0.7} ~keV$ vs $k\rm T=9.8^{+1.0}_{-0.8} ~keV$).
\subsection{MS\,1054.4$-$0321}

MS\,1054$-$0321 is the highest redshift (z=0.83) cluster in the EMSS
and shows a significant amount of substructure with the {\it Chandra}
resolution \citep{Je01}.  As mentioned in Sect.~\ref{comparison} the
temperature and luminosity estimates obtained by ET04 disagree with
ours (see right panels of Fig.~\ref{fig1} and Fig.~\ref{fig4},
respectively).  ET04 estimated both the temperature and the best-fit
surface brightness profile from the main body of the cluster after
masking with a 36$''$ radius circle the cooler region at RA, Dec
(2000)$ = 10^{\rm h} 56^{\rm m} 55^{\rm s}_{.}7, -03^{\circ } 37'
37''$. Since we did not exclude this region a lower temperature and a
higher luminosity are obviously estimated. In addition, this cluster
is an example of how the X-ray temperatures based on {\em Chandra}
data change as new calibrations become available.  An analysis of
XMM-Newton data by \cite{Gi04} results in a temperature $k\rm
T=7.2^{+0.7}_{-0.6} ~keV$, which is much lower than the temperature
previously reported from {\em ASCA} data, $k\rm T=12.3^{+3.1}_{-2.2}
~keV$ \citep{Do98}, and also somewhat lower than the first {\em
Chandra} temperature, $k\rm T=10.4^{+1.7}_{-1.5} ~keV$, determined by
\cite{Je01}.  The temperature measurement of MS\,1054$-$0321 by
\cite{Je01} probably suffered from the absence of a low-energy
correction, called ACISABS, which was not available at the time of
their analysis.  VI02 used the same {\em Chandra} observations and
derived a lower value for the temperature, $k\rm T=7.8\pm0.6 ~keV$, in
agreement with us and with the determination by \cite{To03} of $k\rm
T=8.0\pm0.5 ~keV$. All quoted uncertainties are at 90\% confidence
level except for the last one which is at 68\%.

\smallskip\noindent 
As described in section \ref{fit} we obtain an estimate for the
temperature of $k\rm T=8.3^{+0.7}_{-0.7} ~keV$ within $\rm
R_{spec}=80''$.  We applied the new available procedure to
correct for the quantum efficiency.  As observed in \cite{Je05} this
new procedure (certainly more accurate) tends to yield a higher
temperature than the ACISABS prescription. \cite{Je05} find a
temperature of $k\rm T=8.9^{+1.0}_{-0.8} ~keV$, within
90$''$. Using the more extended region ($R_{ext}$= 130$''$) we
estimate a temperature of $k\rm T=7.8^{+0.9}_{-1.0} ~keV$.
\subsection {MS\,1137.5$+$6625}

MS\,1137.5$+$6625 is the second most distant cluster in the EMSS
sample. Our best-fit temperature of $k\rm T=6.2^{+0.4}_{-0.5} ~keV$ is
consistent with the estimates obtained by VI02 ($k\rm
T=6.3^{+0.4}_{-0.4} ~keV$) and by \cite{Bor01} ($k\rm
T=5.7^{+0.8}_{-0.7} ~keV$). The three temperatures above are computed
within an aperture radius of about 60$''$. Our temperature is also
consistent with the one determined from {\em ASCA} data, $k\rm
T=5.7^{+0.8}_{-0.7}~keV$, by \cite{Do99}.  \cite{Et04} and \cite{To01}
found $k\rm T=6.9^{+0.5}_{-0.5} ~keV$ and $k\rm T=7.0^{+0.5}_{-0.5}
~keV$, respectively, within a smaller region of about 50$''$ radius.

\subsection {RDCSJ\,1317$+$2911}

Despite this cluster is classified by VI02 as a possible cooling core
system, our temperature estimate ($k \rm T=3.7^{+1.2}_{-0.8}$ ~keV) is
higher than the value found by VI02 ($k \rm T=2.2^{+0.5}_{-0.5}$ ~keV)
even though we did not exclude the cooling flow region in our
analysis.  This discrepancy, visible in Fig.~\ref{fig1} (left
panel), might be explained considering that RDCSJ\,1317$+$2911 has a
low signal to noise ratio. A discrepancy in the same direction has
been found by ET04 ($k \rm T = 4.1^{+1.2}_{-0.8}$ ~keV) and by
\cite{To03} ($k \rm T = 4.0^{+1.3}_{-0.8} ~keV$). \cite{To03} argue
that such a difference can be ascribed to differences in the 
procedure used to remove faint point sources within the extraction region, 
which becomes critical for clusters with low S/N such as this one.  This
cluster illustrates the relevance of point source subtraction when
dealing with low number counts.  In fact the point sources increase by
about 60\% the best-fit temperature of RDCSJ\,1317$+$2911 as one can
see in Fig.~\ref{fig5}.  The low S/N implies also large errors on the
luminosity and justifies the disagreement between our results and ET04
shown in Fig.~\ref{fig4}.  The uncertainties on the luminosity are
much larger than the difference expected in the luminosity when using
different radii.
\subsection {ZW\,CL\,1454.8$+$2233}

This cluster is very discrepant with respect to the behavior of other
clusters in the L$_{bol}$--T relationship given in Sect.~\ref{Lbol}.
ZW\,CL\,1454.8$+$2233 was identified as a relaxed cluster hosting a
massive cooling flow by \cite{Al96} using {\em ASCA} and {\em ROSAT}
data.  A 10 Ks {\em Chandra} observation revealed the presence of two
surface brightness edges on opposite sides of the X-ray peak which
were discussed by \cite{Ma01} under the hypothesis of a merging
scenario. The 90 Ks {\em Chandra} observation analyzed by us
confirms a very disturbed morphology. The surface brightness profile 
is inadequately described by a $\beta$-model (the probability to
accept the spatial fit is lower then 0.1\%) both for the presence of 
the cooling core and for the presence of some surface brightness jumps. 
For this reason no values for $R_{c}$ and $\beta$ are indicated in 
Table~\ref{tab2}. On the other hand for the spectral analysis a 
single-temperature model was accepted. The null-hypothesis probability 
is about 5\%, a little less than  the threshold indicated in 
Sect.~\ref{fit}. We tried also to use a cooling flow spectral model 
{\em mkcflow} added to a {\em mekal} model but the improvement in the
fit is minimal.
The thermal complexity of this cluster, which could explain the
peculiarity of its behavior in the L$_{bol}$--T relation, convinced us
to exclude it from the fit of the L$_{bol}$--T relation.

\subsection {MS\,1621.5$+$2640}

The best-fit temperature that we obtained for MS\,1621$+$2640 ($k\rm
T=7.5^{+1.1}_{-0.7} ~keV$) is 10\% greater than the temperature
obtained by ET04 ($k\rm T=6.8^{+0.9}_{-0.5} ~keV$) but in better
agreement with the estimate by VI02 ($k\rm T=7.6^{+0.9}_{-0.9} ~keV$).
Also our luminosity extrapolated to $\rm R_{500}$ disagree with the
estimate of ET04 by an amount on order of 18\%.  This discrepancy
could be partly accounted for by the fact that we found a temperature
of the gas higher with respect to ET04 temperature.
\subsection {CL\,1641$+$4001}

For this cluster the effect of point sources on the cluster thermal
emission is particularly important. In the region covered by the
cluster we found six sources, two of which with a total flux $\geq
50\times10^{-14}\funits$. When the point sources are included in the
fit the luminosity increases of $\sim 50\%$.  On the other hand these
sources have a small ($\sim 3\%$) impact on the cluster
temperature.
 
\subsection {MS\,2053.7$-$0449}

MS\,2053.7$-$0449 is assumed to be a cooling core cluster by
VI02. They measure a temperature of $k\rm T=5.2^{+0.7}_{-0.7} ~keV$.
Our best-fit temperature is $k\rm T=4.3^{+0.5}_{-0.4} ~keV$ within the
$\rm R_{spec}$ region, and $k\rm T=5.1^{+1.4}_{-1.0} ~keV$ when the
larger radius $\rm R_{ext}$ is used. Our lower temperature estimate
could be explained by fact that the possible cooling region is not
excluded in our analysis.  ET04 did not exclude the cooling core and
found a temperature of $k\rm T=5.5^{+0.5}_{-0.5} ~keV$ larger than
VI02 and our temperature estimate. However, \cite{Ma07} found
a value for the temperature of $k\rm T=4.1^{+0.5}_{-0.4} ~keV$ or 
$k\rm T=4.2^{+0.9}_{-0.6} ~keV$ according if the central
region is taken into account or not. From the comparison with the
results of the above mentioned authors it seems that the cluster may
not be a cooling core.

\section{Point source contribution to cluster thermal emission}
\label{cl:ps_cont1}

\begin{figure}
\begin{center}
\resizebox{\textwidth}{!}
{\includegraphics{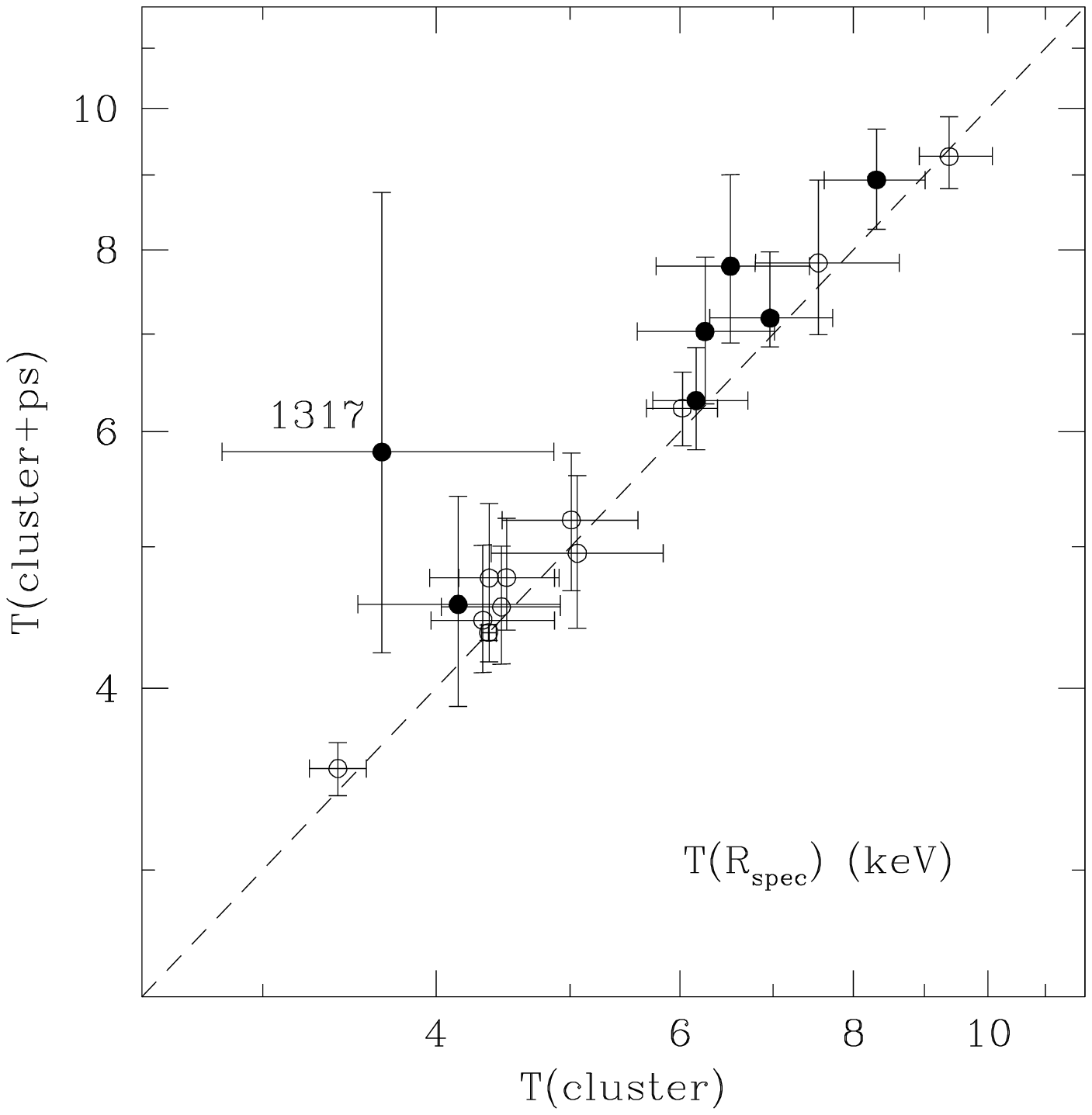}
\includegraphics{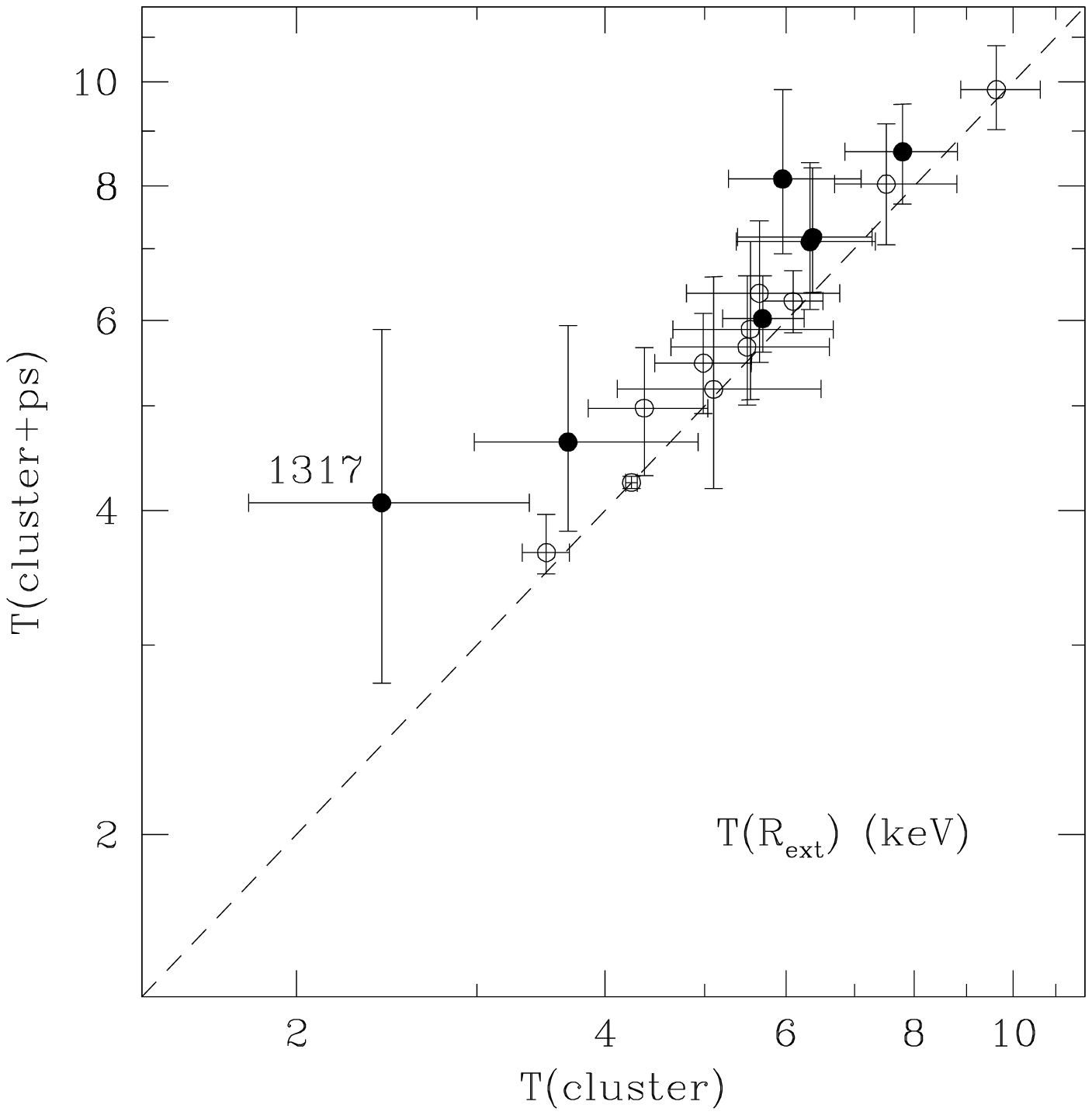}}
\end{center}
\caption{Temperature of clusters plus point sources versus temperature
of clusters without point sources. \textit{Left panel} refers to $\rm
R_{spec}$ and \textit{Right panel} to $\rm R_{ext}$. Solid circles
indicate high-z (z $>$ 0.7) clusters and open circles indicate low-z
(z $<$ 0.7) clusters.  In both panels the dashed line is equality 
between the two temperatures.}
\label{fig5}
\end{figure}

The effect of the point sources on the best-fit temperature is shown
in Fig.~\ref{fig5}. The plots show the estimated temperature for
clusters plus point sources ($\rm T_{cluster+ps}$) versus the
temperature for clusters without point sources ($\rm T_{cluster}$).
The left panel refers to temperatures estimated within $\rm R_{spec}$
while the right panel refers to $\rm R_{ext}$. 
Obviously the number of point sources within $\rm R_{ext}$ is higher than 
within $\rm R_{spec}$. The most discrepant point in Fig.~\ref{fig5} has 
been discussed in the note of RDCSJ\,1317$+$2911 (Sect.~\ref{notes}).

\smallskip\noindent
The effect of the point sources on the temperature estimate is evident
(especially within $\rm R_{ext}$). In addition a mild dependence 
on the redshift is also present. The average of the ratio 
$\rm r_T = T_{cluster+ps}/T_{clusters}$ gives a temperature
excess of $(8\pm 3)$\% within $\rm R_{spec}$ and $(13\pm 4)$\% within
$\rm R_{ext}$. Splitting the sample at z $=0.7$ we get, respectively,
$(16\pm 7)$\% and $(24\pm 8)$\% for z $>$ 0.7 and $(3\pm 1)$\% and
$(6\pm 1)$\% for z $<0.7$.

\begin{figure}
\begin{center}
\resizebox{\textwidth}{!}
{\includegraphics{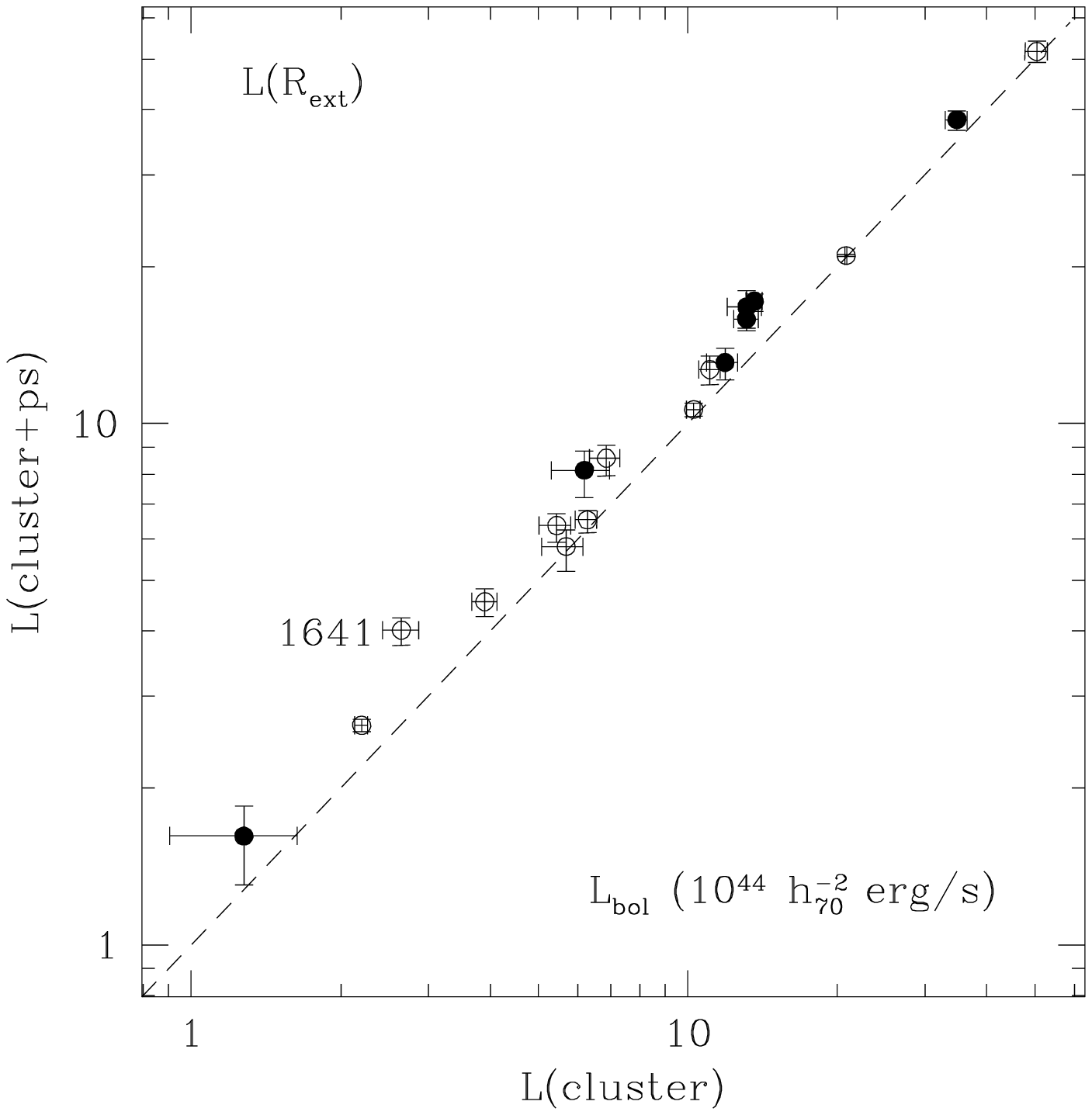}\includegraphics{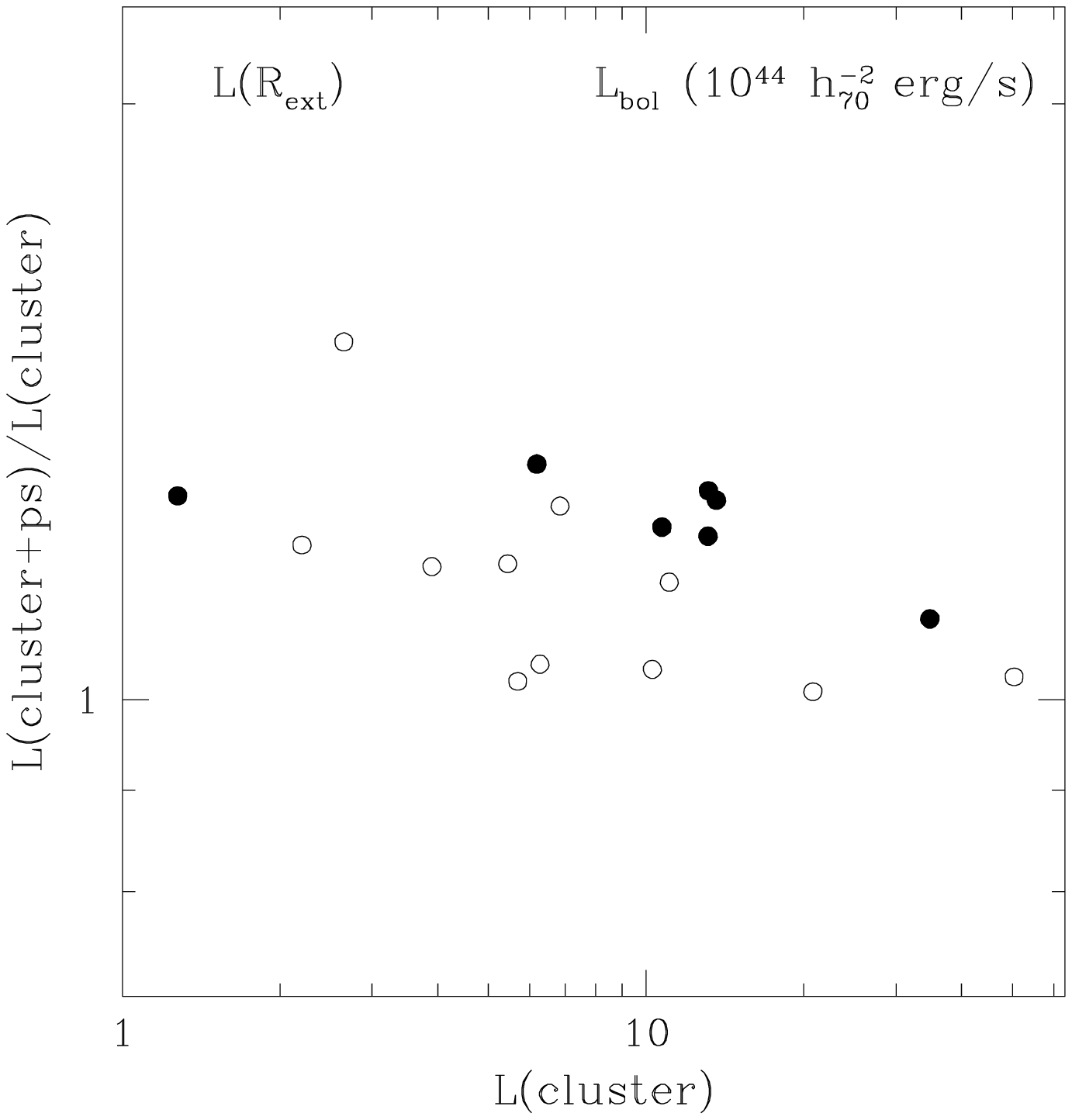}}
\end{center}
\caption{\textit{(Left panel)} Luminosity of clusters plus point
sources versus luminosity of clusters without point sources. The dashed 
line is equality between the two luminosities. \textit{(Right 
panel)} Ratio between luminosity of the clusters plus point sources
and luminosity of the clusters without point sources versus  luminosity of 
the clusters alone. Solid circles refer to high-z clusters and open circles 
refer low-z clusters.}
\label{fig6}
\end{figure}

\smallskip\noindent 
The left panel of Figure~\ref{fig6} represents the same plot of
Fig.~\ref{fig5} for cluster luminosities: $\rm L_{cluster+ps}$ is
plotted versus $\rm L_{cluster}$.  A different way of presenting these
data is illustrated to the right of Fig.~\ref{fig6} where the ratio
$\rm r_L = L_{cluster+ps}/L_{cluster}\approx (1+L_{ps}/L_{cluster})$
is plotted versus $\rm L_{cluster}$. This is a way to examine the
``contrast'' of point source luminosity with respect to cluster
luminosity. The plot shows a mild trend for the luminosity
``contrast'' to decrease with increasing cluster luminosity.
Similarly to the temperature, the contrast in luminosity due to the
point sources is more pronounced in distant clusters. The low-z
clusters (open circles in Fig.~\ref{fig6}, right) are generally below
the high-z clusters (solid circles) in the same luminosity range. The
average of $\rm r_L$ gives an over-luminosity of $(17\pm 3)$\% with
$(22\pm 3)$\% for high-z clusters and of $(14\pm 4)$\% for low-z
clusters.  These values are similar, but more significant, than those
for $\rm r_T$ computed within $\rm R_{ext}$.

%

\begin{figure}
\begin{center}
\resizebox{\textwidth}{!}{\includegraphics{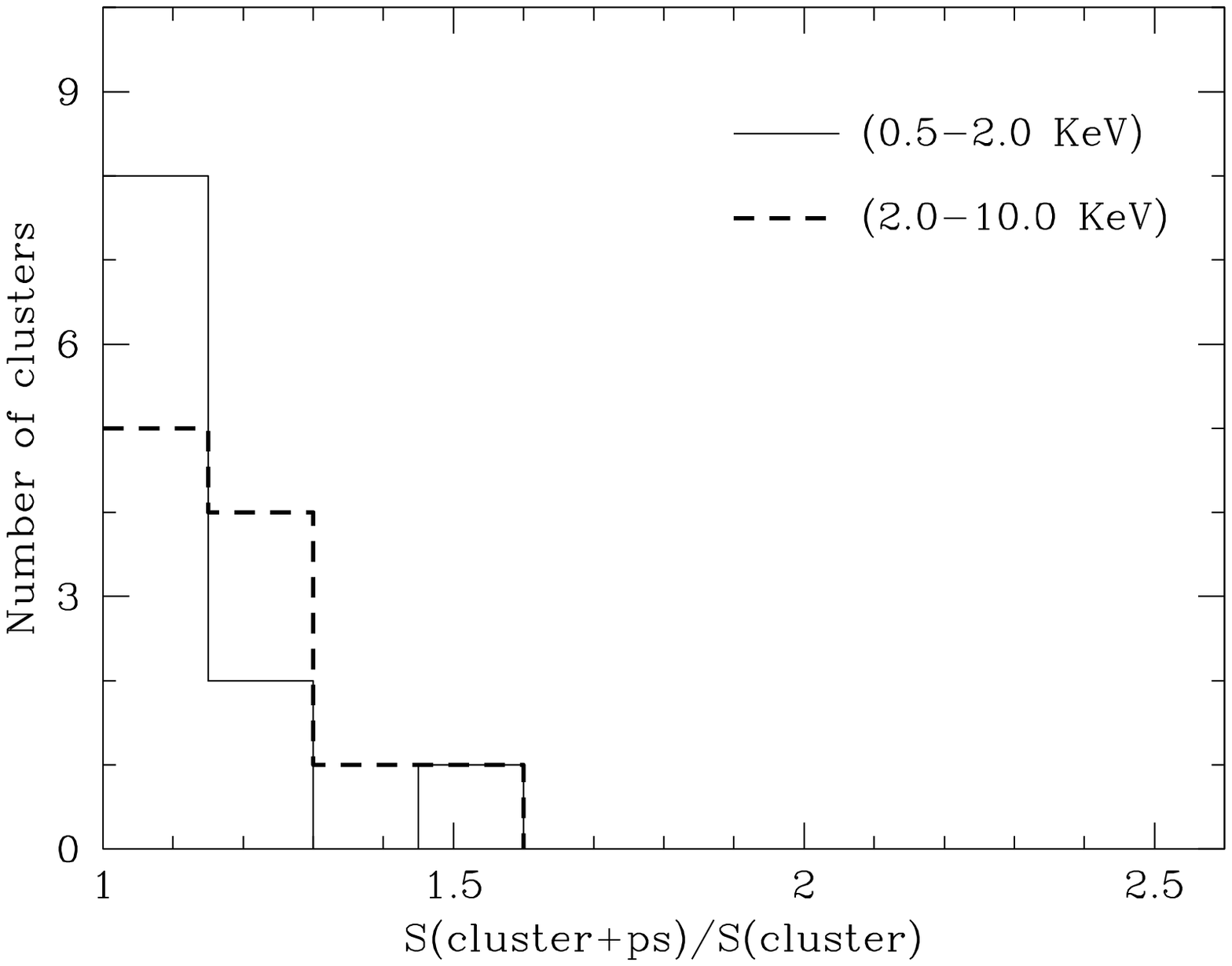}
\includegraphics{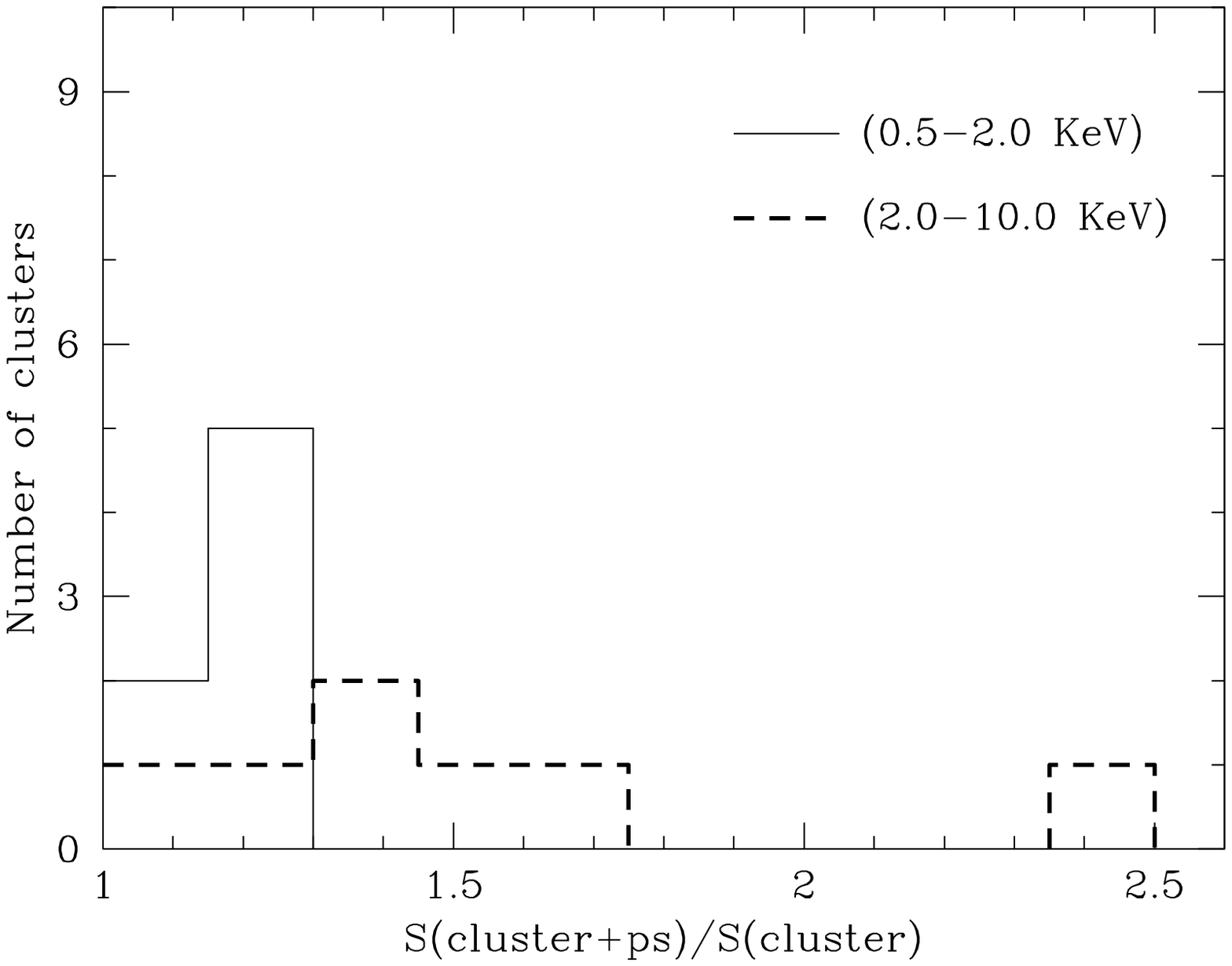}}
\end{center}
\caption{Histograms of the ratio between the flux of the clusters plus
point sources and the flux of clusters without point sources. The
\textit{left panel} is for clusters with redshift $<$ 0.7 and the
\textit{right panel} for clusters with redshift$>$ 0.7. The solid
line indicates the soft energy band (0.5--2.0 keV) and the dashed line
indicates the hard energy band (2.0--10.0 keV).}
\label{fig7}
\end{figure}

\medskip\noindent
Given the probably different X-ray spectrum of clusters and point
sources (possibly AGN) we have investigated the contribution of the
point sources to the flux in the soft and hard energy bands
separately, by computing the two ratios $\rm r_S$ = $\rm
S_{cluster+ps}/S_{cluster}$ for both 0.5--2.0 keV and 2.0--10.0 keV
energy bands.  The histograms of such ratios are shown, overplotted,
in Fig.~\ref{fig7}, where the solid (dashed) line represents the
soft (hard) band. Clusters with redshift $<$ 0.7 are shown to the left 
and those with redshift $>$ 0.7 to the right.  The figure indicates that 
the point source contribution is higher in the hard band, as one might 
expect if the point sources have a spectrum harder than clusters.  
This effect is more evident for the distant clusters (right panel). It has
to be noted that in both energy bands the excess in flux is mainly due
to some bright objects (projected onto or belonging to the cluster)
rather than to many faint sources.


\smallskip\noindent
The data suggest that there is a mild tendency for distant clusters (z
$>0.7$) to be more affected by point sources.  If real, this effect
would naively imply that the line of sight to farther clusters
intercepts a higher number of bright point sources. However, this is
contrary to what expected from the fact that distant clusters cover an
average angular area which is $\sim$~50\% that covered by nearby
clusters, and that the flux is about half the flux of nearby clusters.
If point sources are mostly unrelated to the clusters, these two
effects would compensate and the point source contribution to the
cluster overall budget should be independent of the cluster
redshift. The observational result of seeing a larger number of point
sources in distant clusters goes in the direction of having more
sources belonging to high-z clusters. Indeed in a recent paper BR07
found an overdensity of point sources in clusters with respect to the
field, and an indication in the hard band that the excess is mainly
associated to high-z cluster.

\section{The L$_{bol}$--T  relationship with and without point sources}
\label{Lbol}

The effect of the point sources on L$_{bol}$--T relation can be
evaluated by direct comparison of the relations obtained by including
and excluding the point sources.  We first analyzed the L$_{bol}$--T
using the gas temperature estimated within $\rm R_{spec}$. Then, in
order to consider the contribution on the temperature of all the
sources within the extended region where the luminosity is computed,
we used the gas temperature estimated within $\rm R_{ext}$.
We express the L$_{bol}$--T relation as
\begin{eqnarray}
{\rm L_{bol,44}} = C {\rm T_{6}^{\alpha}}
\end{eqnarray}
\noindent
where $\rm L_{bol,44}$ is the bolometric luminosity in units of $10^{44}$ erg 
s$^{-1}$ and $\rm T_6 = T(keV)/6$.

\begin{figure}
\begin{center}
\resizebox{\textwidth}{!}{\includegraphics{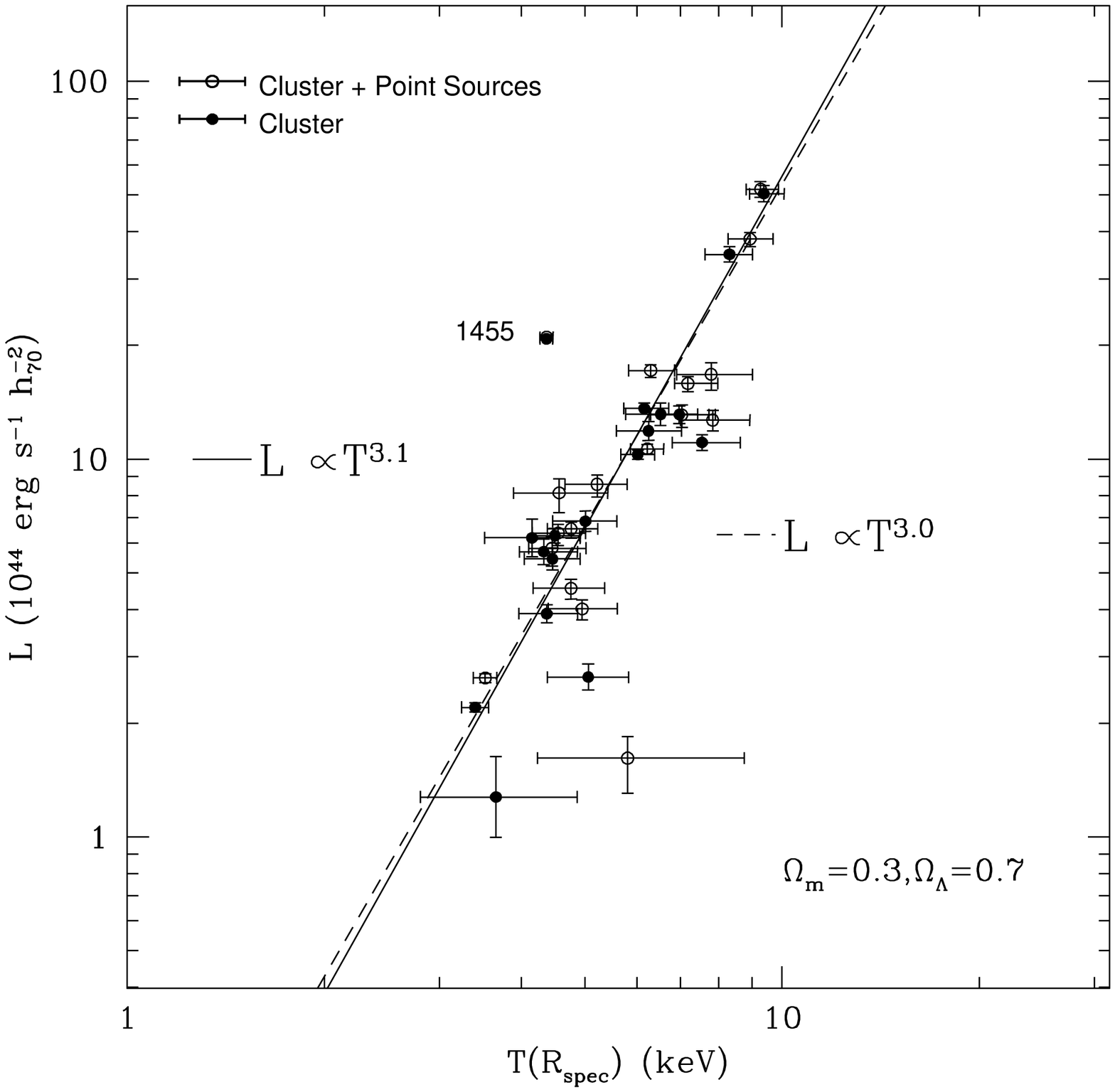}
\includegraphics{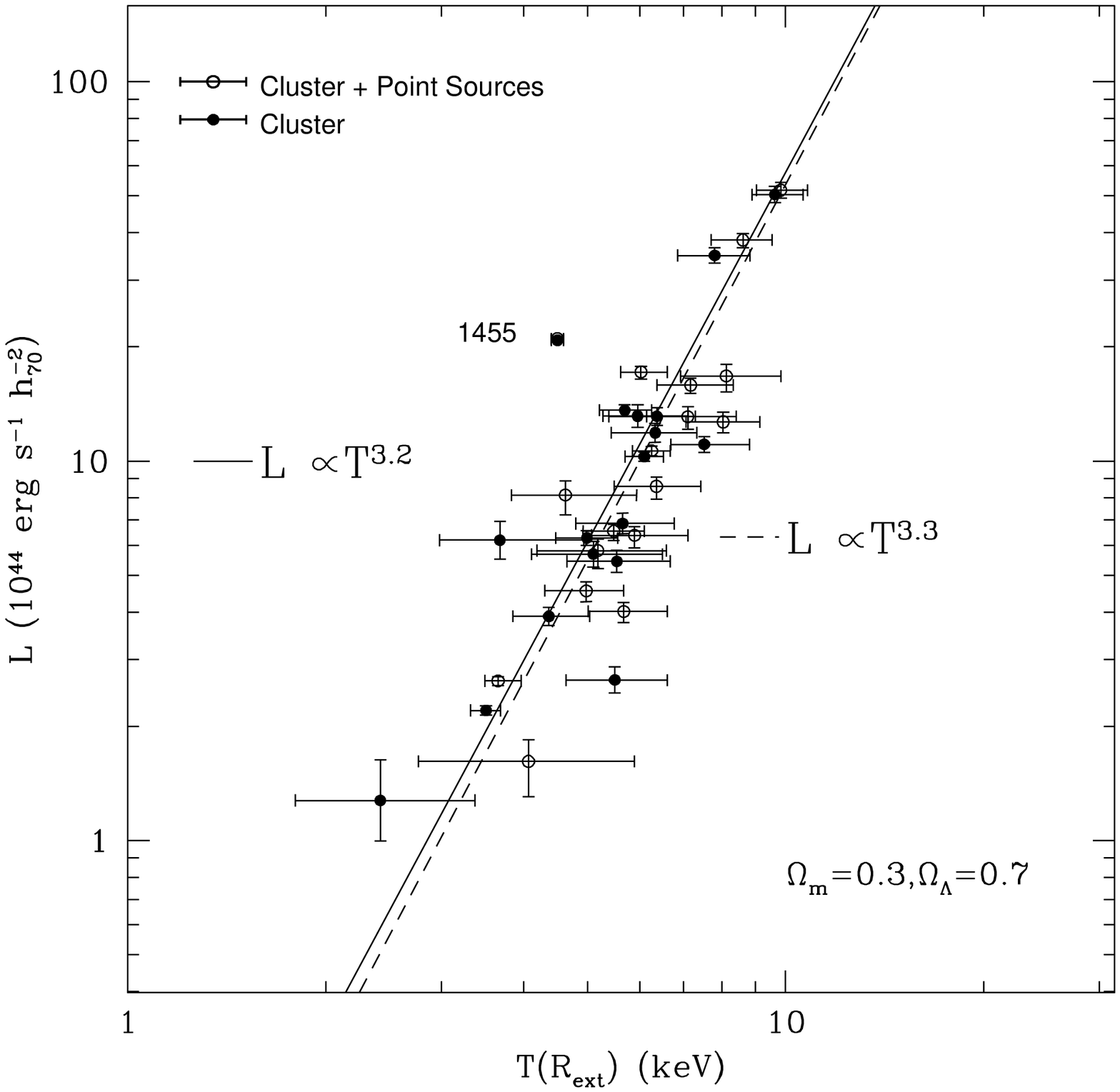}}
\end{center}
\caption{Cluster bolometric luminosity versus temperature
relationship. The open circles indicate cluster plus point source
emission and the solid circles indicate cluster emission alone. The
dashed and solid lines are the best-fit of the L$_{bol}$--T relation
to the open and solid circles data respectively.  The  solid
circle indicated as 1455 represents the cluster ZW\,CL\,1454.8$+$2233
which was removed from the fit (see Notes on individual clusters in
Sect.~\ref{notes}).}
\label{fig8}
\end{figure}

\smallskip\noindent
The data and the fitted L$_{bol}$--T (using a $\chi^{2}$ method which 
takes into account the $\rm L_{bol,44}$ and T uncertainties) are
displayed in Fig.~\ref{fig8} for both the $\rm R_{spec}$ (left) and
$\rm R_{ext}$ (right), and the best-fit parameters are listed in
Table~\ref{tab3}.  Solid and open symbols and solid and dashed lines
represent quantities with and without point source inclusion. The most
discrepant point in both plots is the cluster ZW\,CL\,1454.8$+$2233
which was excluded from the fit of the L$_{bol}$--T relation (see
Notes on individual clusters in Sect.~\ref{notes}).
%
The inclusion of the point sources, for both ${\rm R_{spec}}$ and
${\rm R_{ext}}$, does not have any significant effect on the slope and
normalization of the L$_{bol}$--T relation. This is because the
correction to T and ${\rm L_{bol}}$ applied to each data point in the
plot produces a shift in the T$-{\rm L_{bol}}$ plane almost parallel
to the best-fit line. In Paper II (Branchesi et al. 2007) we analyze
in more detail the L$_{bol}$--T relation using data free of point
sources.
\begin{table}
\begin{center}
\caption{$\rm L_{bol}-T$ best-fit parameters}
\begin{tabular}{c|cccc}
\hline
\hline
$\rm L_{bol}-T(R_{spec})$   & Cluster & Cluster + Point Sources\\
\hline
\hline
$\alpha$  & $+3.09^{~+~0.35}_{~-~0.28}$  &  $+3.00^{~+~0.31}_{~-~0.26}$\\
\\
log C &  $+1.06^{~+~0.04}_{~-~0.04}$ & $+1.06^{~+~0.04}_{~-~0.04}$ \\
\\
$\chi_{min}^2/d.o.f.$ &  14.77/15  & 17.54/15 &\\
\hline
\\
$\rm L_{bol}-T(R_{ext})$   & Cluster & Cluster + Point Sources\\
\hline
\hline
$\alpha$  & $+3.23^{~+~0.51}_{~-~0.38}$  &  $+3.29^{~+~0.58}_{~-~0.43}$ \\
\\
log C & $+1.04^{~+~0.07}_{~-~0.06}$ & $+1.00^{~+~0.06}_{~-~0.06}$ \\
\\
$\chi_{min}^2/d.o.f.$ & 12.24/15  & 14.25/15  &\\
\label{tab3}
\end{tabular}
\end{center} 
\end{table}

\section{Summary and Conclusions}

In this paper we have presented the details of the data analysis of 
a sample of 18 distant clusters (0.25 $<$ z $<$ 1.01) taken from the 
{\it Chandra} archive to derive the observational properties of the 
X-ray emitting gas. The same sample was used to study the point source 
counts in the inner region of distant clusters (BR07) and to study the 
evolution with redshift of the L$_{bol}$--T relation  (Paper II, Branchesi 
et al. 2007).

\smallskip\noindent
The very high angular resolution of {\em Chandra} that allows to
isolate X-ray point sources embedded in the more extended X-ray
emission from galaxy clusters, enabled us to estimate for the 18
clusters the effect of the point source non-thermal emission on the
determination of the thermal emission due to the cluster itself.  

\smallskip\noindent
{\it i)} The point sources located within the cluster thermal
emission region may affect considerably the estimates of X--ray 
observables like cluster temperature (by an amount up to 13\%) and 
luminosity (by an amount of 17\%). These percentages become larger if 
one considers clusters with z$>$0.7 where temperature and luminosity 
increase up to 24\% and 22\%, respectively (see Section \ref{cl:ps_cont1}).  
The results obtained suggest that, in order to estimate properly the 
observational parameters of the thermal emission, the point source 
contribution should be removed. 

\smallskip\noindent
{\it ii)} However, the inclusion (or exclusion) of point sources in the
analysis of the L$_{bol}$--~T relation, indicates minor differences,
within the uncertainties, of the relation (see Section \ref{Lbol}). 
This is due to the fact that for each cluster the correction to be 
applied to T and L$_{bol}$ for the presence of point sources produces 
a moderate shift in the L$_{bol}$--T plane almost parallel to the best-fit 
of the ``correct'' (excluding the point sources) L$_{bol}$--T relation.
\begin{acknowledgements}
This research made use of data obtained from the Chandra Data Archive, 
which is part of the Chandra X-Ray Observatory Science Center, operated 
for the National Aeronautics and Space Administration (NASA) by the 
Smithsonian Astrophysical Observatory. Partial financial 
support for this work came from the Italian Space Agency ASI (Agenzia 
Spaziale Italiana) through grant ASI-INAF I/023/05/0.
\end{acknowledgements}

\clearpage
\appendix
\section{Analysis of the best-fit parameters}
\subsection{Comparison between Cash and $\chi^2$ statistics}
\label{appendix:a1}

\begin{figure}
\begin{center}
\resizebox{\textwidth}{!}
{\includegraphics{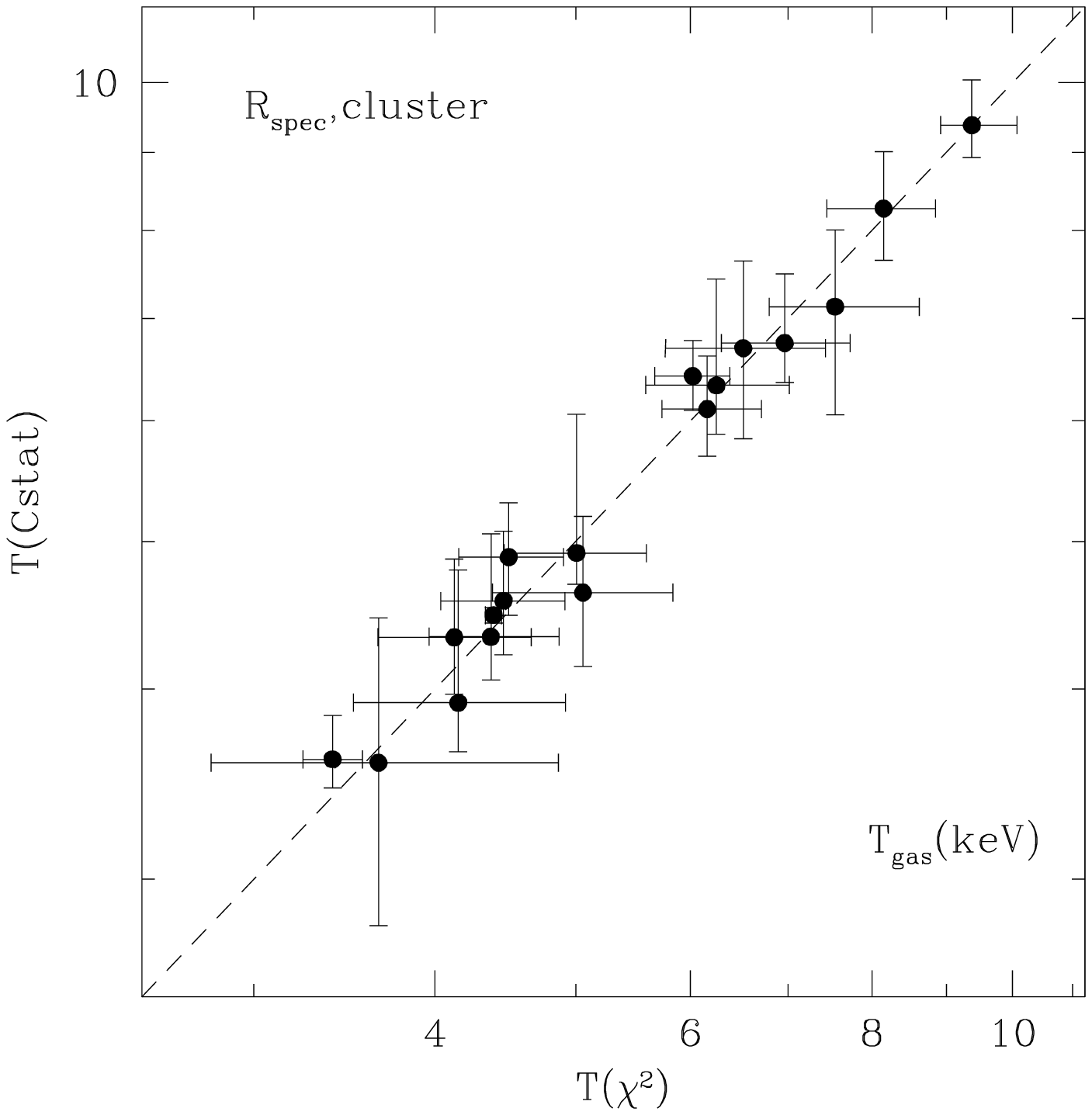}\includegraphics{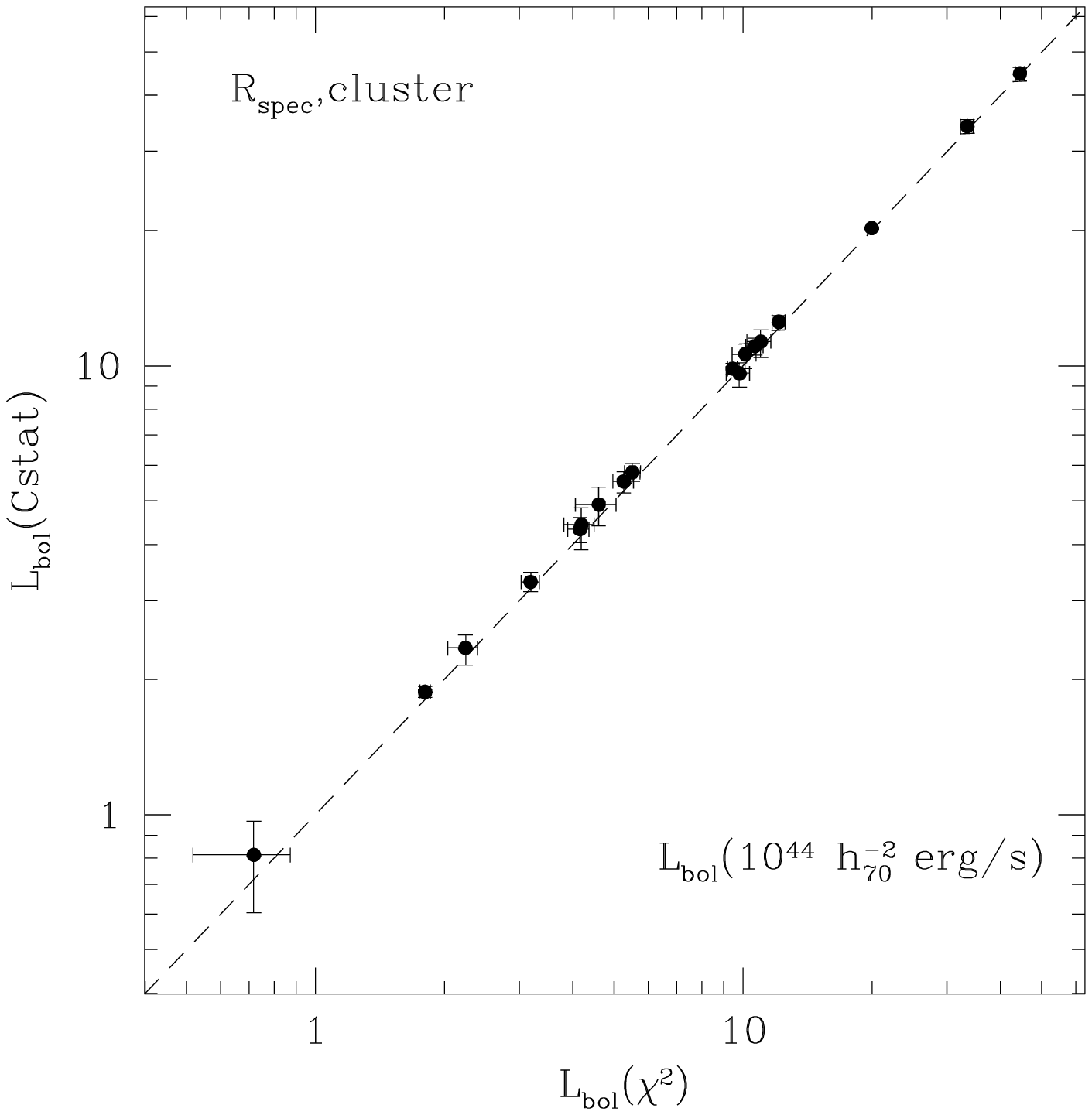}}
\end{center}
\caption{(\textit{Left panel}) The best-fit temperature obtained using 
the Cash statistics versus the best-fit temperature obtained with the 
$\chi^2$ statistics. The dashed line is equality between the two temperatures.
(\textit{Right panel}) The luminosity obtained using the Cash statistics 
versus the best-fit luminosity obtained using the $\chi^2$ statistics. The
dashed line is equality  between the two luminosities.}
\label{appendix:fig1}
\end{figure}

In this Appendix we compare the best-fit temperatures and the
bolometric luminosities obtained with the Cash and the $\chi^2$
statistics only for those spectrum files with the point sources
excised.  The agreement is remarkably good when the spectra are
extracted within the region defined using $\rm R_{spec}$ (see
Fig.~\ref{appendix:fig1}). The temperatures (luminosities) obtained
with the Cash statistics are, on average, $\sim 1\%$ ($\sim 4\%$)
higher then those obtained with the $\chi^2$ statistics.  The
histogram of the ratios between the temperatures (luminosities)
obtained with the $\chi^2$ and the Cash statistics is characterized by
a standard deviation of 0.04 (0.03).

\smallskip\noindent 
The agreement remains good considering the more extended region
defined when one uses the radius $\rm R_{ext}$. In this case the
luminosities with the Cash statistics are on average $\sim 3\%$
smaller than those obtained with $\chi^2$, while the temperatures
remain the same.  The dispersion of the temperature histogram obtained
within the region defined by $\rm R_{ext}$ has a standard deviation of
0.1, higher than the dispersion obtained when using $\rm R_{spec}$.
This is presumably due to the fact that the spectral fit is better in
the region that optimizes the signal to noise ratio.

\subsection{Best Parameters for Luminosity and Temperature} 
\label{appendix:a2}

\begin{figure}
\includegraphics[width=12cm,height=12cm,bb=-150 50 424 624]{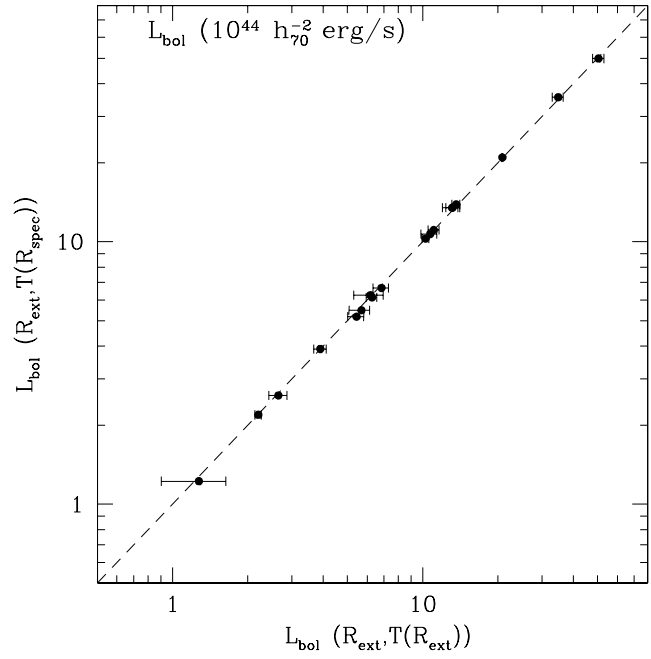}
\vspace{-2cm}
\caption{Best-fit luminosity obtained within $\rm R_{ext}$ with the temperature 
fixed to the value obtained within $\rm R_{spec}$ versus best-fit luminosity 
obtained within $\rm R_{ext}$. The dashed line is equality between the two 
best-fit luminosities.}
\label{appendix:fig2}
\end{figure}

The quality of the fits is better inside the region within $\rm
R_{spec}$, e.g the region that maximizes the signal to noise
ratio. Thus the temperatures estimated within this region, $\rm
T_{R_{spec}}$, are considered more representative of the actual
average temperature of the gas.  On the other hand the cluster
luminosities are estimated within the extended region defined by the
radius $\rm R_{ext}$ in order to take into account the faint
brightness tails at the cluster boundaries as well as the point
sources in those regions. For each cluster the luminosity was
calculated by fitting the counts accumulated within $\rm R_{ext}$ with
a thermal bremsstrahlung model.  A best-fit temperature, indicated as
$\rm T_{R_{ext}}$, is associated to each thermal model. Systematic
differences between {$\rm T_{R_{spec}}$} and {$\rm T_{R_{ext}}$}
are within 3\%.
\smallskip\noindent 
In order to understand how strongly the luminosity depends on the
model and on its associated temperature, and in order to know which
errors are involved when temperature and luminosity are calculated in
two different regions (i.e. defined by $\rm R_{spec}$ and $\rm
R_{ext}$, respectively) we computed the luminosity of each cluster in
the $\rm R_{ext}$ region but using the temperature obtained within
$\rm R_{spec}$. The results of this exercise are shown to the left of
Fig.~\ref{appendix:fig2}.  The two methods give values consistent
within the errors.  The differences between the luminosities are all
within 5\%, that is within the statistical errors. 

\noindent
\end{document}